\documentclass[twocolumn,journal]{IEEEtran}
\usepackage[T1]{fontenc}
\usepackage[latin9]{inputenc}
\usepackage{amsmath}
\usepackage{amssymb}
\usepackage{stackrel}
\usepackage{graphicx}
\usepackage[unicode=true,
 bookmarks=true,bookmarksnumbered=true,bookmarksopen=true,bookmarksopenlevel=1,
 breaklinks=false,pdfborder={0 0 0},pdfborderstyle={},backref=false,colorlinks=false]
 {hyperref}
\hypersetup{pdftitle={Your Title},
 pdfauthor={Your Name},
 pdfpagelayout=OneColumn, pdfnewwindow=true, pdfstartview=XYZ, plainpages=false}

\makeatletter
\usepackage[caption=false,font=footnotesize]{subfig}

\@ifundefined{showcaptionsetup}{}{%
 \PassOptionsToPackage{caption=false}{subfig}}
\usepackage{subfig}
\makeatother

\begin{document}
\title{Enhancing Ambient Backscatter Communication Utilizing Coherent and
Non-Coherent \\
Space-Time Codes}
\author{Wenjing~Liu,~\IEEEmembership{Student Member,~IEEE,}~Shanpu~Shen,~\IEEEmembership{Member,~IEEE,}
Danny~H.~K.~Tsang,~\IEEEmembership{Fellow,~IEEE,}~Ross~Murch,~\IEEEmembership{Fellow,~IEEE}\thanks{Manuscript received; This work was supported by the Hong Kong Research
Grants Council under General Research Fund grant 16207620. \textit{(Corresponding
author: Shanpu Shen.)}}\thanks{W. Liu, S. Shen, D. H. K. Tsang and R. Murch are with the Department
of Electronic and Computer Engineering, The Hong Kong University of
Science and Technology, Clear Water Bay, Kowloon, Hong Kong (e-mail:
wliubj@connect.ust.hk; sshenaa@connect.ust.hk; eetsang@ust.hk; eermurch@ust.hk).}}
\maketitle
\begin{abstract}
Ambient backscatter communication (AmBC) leverages the existing ambient
radio frequency (RF) environment to implement communication with battery-free
devices. The key challenge in the development of AmBC is the very
weak RF signals backscattered by the AmBC Tag. To overcome this challenge,
we propose the use of orthogonal space-time block codes (OSTBC) by
incorporating multiple antennas at the Tag as well as at the Reader.
Our approach considers both coherent and non-coherent OSTBC so that
systems with and without channel state information can be considered.
To allow the application of OSTBC, we develop an approximate linearized
and normalized multiple-input multiple-output (MIMO) channel model
for the AmBC system. This MIMO channel model is shown to be accurate
for a wide range of useful operating conditions. Two coherent detectors
and a non-coherent detector are also provided based on the proposed
AmBC channel model. Simulation results show that enhanced bit error
rate performance can be achieved, demonstrating the benefit of using
multiple antennas at the Tag as well as the Reader.
\end{abstract}


\section{Introduction}

\IEEEPARstart{B}{ackscatter} communication has a long history \cite{reflected_1948_stockman},
\cite{history_rfid_Landt_2005} and an important application is Radio
Frequency (RF) Identification. Typically in backscatter communication
systems, a ``Reader'' transmits signals to a ``Tag'', which harvests
energy from the received signals to power its circuits and backscatter
the signals to the Reader through tuning its load impedance \cite{bistatic_Kimionis_2013}-\nocite{MIMO_RFID_Boyer_2014}\cite{survery_VanHuynh_2018}.
The key advantage of backscatter communication is that the Tag does
not require a battery and can be inexpensive to construct. However,
there are some disadvantages that limit the use of backscatter communication.
These include the need for a dedicated source of RF radiation as well
as a frequency spectrum allocation to allow the system to operate.

To address the disadvantages of conventional backscatter communication,
ambient backscatter communication (AmBC) systems have been proposed
\cite{thin_Liu_2013}. Compared with traditional backscatter systems,
AmBC harnesses ambient RF signals transmitted from existing wireless
systems (such as DTV \cite{thin_Liu_2013}, FM \cite{FMbackscatter_Wang_2017},
\cite{AmBC_FM_Daskalakis_2017}, and Wi-Fi \cite{WiFiback_Kwllogg_2014}-\nocite{Backfi_Bharadia_2015}\nocite{passiveWIFI_Kellogg_2016}\cite{Efficient_BackFi_Ji_2019})
as sources of RF radiation. This allows AmBC to operate without any
dedicated RF source or extra frequency spectrum allocation and makes
it a promising technology for applications such as the Internet-of-Things
(IoT).

A significant challenge of AmBC is that the ambient RF waves are unknown
and uncontrollable and the backscattered signals are often very weak.
To solve this issue, an averaging mechanism has been proposed \cite{thin_Liu_2013}
to eliminate the influence of the ambient signals. Based on this idea,
energy detection with detection thresholds and bit error rate (BER)
expressions have been developed \cite{SISO_Lu_2015}, \cite{Exact_BER_Devineni_2019}.
Differential encoding and non-coherent energy detectors have also
been designed, which do not require any channel state information
(CSI) and pilot symbols \cite{Detection_perfprmance_Wang_2016,Nocoherent_Qian_2017}.
The non-coherent energy detector has also been generalized to MPSK
\cite{MPSK_Qian_2019}. Instead of averaging the ambient RF signals,
FSK \cite{MFSK_Tao_2019}, \cite{Swiching_frequency_Vougioukas_2019}
has been proposed so that the backscattered signals shift to a band
that does not overlap the ambient RF signals. In \cite{matched_filter_Choi_2019},
matched-filtering at the Tag is utilized to estimate ambient waves
as well as detect backscattered signals. Coding mechanisms have also
been utilized in AmBC systems, including a three states coding scheme
\cite{Coding_Detection_Liu_2017} to improve throughput as well as
Manchester coding \cite{Manchester_coding_Tao_2018} to reduce the
decoding delay. Moreover, performance analysis in terms of resource
allocation and scheduling have been proposed \cite{Optimal_Resource_Allocation_Full_Duplex_Yang_2019}-\nocite{Time_scheduling_Liu_2019}\nocite{TP_Splitting_Ma_2019}\cite{Modeling_performance_Cellular_Shi_2020}.
In \cite{Time_scheduling_Liu_2019}, \cite{TP_Splitting_Ma_2019},
time switching and power splitting architectures have been optimized
to achieve optimal outage and throughput performance. Aspects related
to ergodic capacity and outage probability have also been studied
\cite{Modeling_performance_Darsena_2017}-\nocite{Ergodic_rat_Zhou_2019}\nocite{outage_cooperative_Ding_2020}\cite{Outage_performance_Ye_2020}.
In \cite{Approximate_BER_Tagselection_Zhou_2017}, \cite{Capacity_Tag_delection_Li_2019},
the BER and capacity of a Tag selection scheme has been explored and
in \cite{Roubust_design_Zhang_2019}, throughput of a multitag AmBC
system has been considered. AmBC has also been incorporated with RF-powered
cognitive ratio networks and the secondary network throughput has
been significantly improved \cite{New_approach_RF_POWER_Hoang_2017},
\cite{Opportunistic_AmBC_2019}. In addition, the concepts of energy
efficiency and hardware efficiency for AmBC have also been introduced
\cite{Opportunistic_AmBC_2019}, \cite{Hardware_Efficient_detection_Tao_2019}.

To further enhance backscattered signal detection performance, multiple
antennas have been proposed at the Reader so that diversity gain can
be leveraged \cite{umo_Parks_2014}-\nocite{expand_umo_Ma_2018}\nocite{eigenvalue_detection_Tao_2019}\nocite{multiple_antenna_cognitive_Guo_2019}\nocite{Constellation_Learning_Based_Zhang_2019}\nocite{frequency_diverse_array_Hu_2020}\nocite{Cooperative_YCLiang_2018}\nocite{MRC_Adaptive_Li_2018}\nocite{OFDM_multiple_receive_Yang_2018}\nocite{multi_antenna_Duan_2018}\cite{Nocohrent_OFDM_ElMossallamy_2019}.
Specifically, a dual-antenna Reader and a ratio detector have been
proposed \cite{umo_Parks_2014} to extend the transmission distance.
Based on these results \cite{umo_Parks_2014}, the ratio detector
has been re-investigated \cite{expand_umo_Ma_2018} and the corresponding
decision threshold when the Reader has more than two antennas has
been found. The problem of noise uncertainty in multiple-antenna Reader
AmBC systems has also been solved \cite{eigenvalue_detection_Tao_2019}.
By using receive beamforming at a multiple-antenna receiver \cite{multiple_antenna_cognitive_Guo_2019},
and by a constellation-learning method \cite{Constellation_Learning_Based_Zhang_2019},
direct-link interference is mitigated and CSI can be avoided. Multi-antenna
frequency diverse arrays have been employed to improve channel capacity
\cite{frequency_diverse_array_Hu_2020}. In addition, a cooperative
AmBC system \cite{Cooperative_YCLiang_2018}, an adaptive AmBC system
\cite{MRC_Adaptive_Li_2018}, and AmBC systems leveraging ambient
orthogonal frequency division multiplexing (OFDM) modulated signals
\cite{OFDM_multiple_receive_Yang_2018}-\nocite{multi_antenna_Duan_2018}\cite{Nocohrent_OFDM_ElMossallamy_2019},
all with multiple-antenna Readers, have been proposed to enhance AmBC
system performance.

Using multiple antennas at the Tag can also provide diversity gain
to enhance signal detection performance, however, there are only a
few studies considering this. Compared with using multiple antennas
at the Reader, multiple antennas at the Tag can increase the backscattered
signal power and the amount of ambient RF energy that can be harvested
\cite{ShanpuShen2017_TAP_EHPIXEL}. It should be noted however that
increasing antenna directivity at the Tag through using RF combining
or large antennas cannot straightforwardly improve backscatter. This
is because the antenna will become directive with the main beam potentially
pointing away from the Reader and reducing backscatter to the Reader.
Therefore, more advanced approaches must be used for multiple-antenna
Tags. An AmBC system with different power allocations on two Tag antennas
has been proposed \cite{Signal_detection_multiple_antenna_Tag_Kang_2017}
to increase link reliability. However, an 8-bit preamble is required
to calculate the decision threshold, and CSI at the Reader (CSIR)
is required. In recent studies \cite{Transceiver_Chen_2020,optimal_antenna_Chen_2020},
a multiple-antenna Tag has been proposed where antenna selection is
performed to increase the diversity gain of the AmBC system, and the
information is recovered utilizing the difference in signal power
statistics. However, the proposed approaches \cite{Transceiver_Chen_2020,optimal_antenna_Chen_2020}
achieve optimal performance based on knowing the order of the channel
gains at the Tag, which means that partial CSI at the Tag (CSIT) is
required. Hence, pilot symbols and feedback from the Reader to Tag
are required, which increases power consumption, time delay, and system
complexity.

\subsection{Motivation and Contributions}

The majority of the work reviewed \cite{SISO_Lu_2015}-\cite{Nocohrent_OFDM_ElMossallamy_2019}
only uses a single antenna at the Tag. Only a few works \cite{Signal_detection_multiple_antenna_Tag_Kang_2017}-\nocite{Transceiver_Chen_2020}\cite{optimal_antenna_Chen_2020}
consider using multiple antennas at the Tag, but these works only
use a single antenna at the Reader and therefore, receive diversity
gain has not been leveraged. In addition the majority of the previous
work using multiple antennas also require CSI knowledge \cite{umo_Parks_2014},
\cite{expand_umo_Ma_2018}, \cite{Cooperative_YCLiang_2018}-\nocite{MRC_Adaptive_Li_2018}\cite{OFDM_multiple_receive_Yang_2018},
\cite{Signal_detection_multiple_antenna_Tag_Kang_2017}-\nocite{Transceiver_Chen_2020}\cite{optimal_antenna_Chen_2020}.
For example, CSIR \cite{Signal_detection_multiple_antenna_Tag_Kang_2017}
and CSIT \cite{Transceiver_Chen_2020}, \cite{optimal_antenna_Chen_2020}.
However, estimating CSI increases the complexity and power consumption
of AmBC.

To the best of our knowledge there is no work using multiple antennas
at both the Tag and Reader that leverage conventional multiple-input
multiple output (MIMO) techniques to enhance signal detection performance
in AmBC systems. Performing this would deliver power gain, transmit
diversity gain, receive diversity gain and enhance signal detection
in AmBC. Methods for providing it with low complexity and without
requiring CSIR or CSIT would also be desirable.

In this paper, we propose using a multiple-antenna Tag and a multiple-antenna
Reader in AmBC by applying coherent and non-coherent orthogonal space-time
block codes (OSTBC) to enhance the detection performance. Specifically
our contributions include:

1) Using multiple antennas both at the Tag and at the Reader: We propose
using multiple antennas at both the Tag and Reader to form a MIMO
AmBC system, so as to jointly leverage power gain, transmit diversity
gain, and receive diversity gain to enhance signal detection in AmBC.
We show through both theoretical analysis and simulation results that
using multiple antennas at both sides can significantly lower the
BER and increase the total harvested energy. To the authors\textquoteright{}
best knowledge, this is the first work which proposes a MIMO AmBC
system to enhance signal detection performance.

2) Proposing a linear MIMO channel model for the multiple antenna
Tag and Reader system: In particular we derive an accurate AmBC channel
model that can be applied to any MIMO AmBC communication scenario.
Due to the MIMO AmBC channels nonlinear nature we develop an approximate
linear MIMO AmBC channel model to simplify the detection process so
that any conventional MIMO technique can be utilized. This is a key
contribution in enabling MIMO AmBC.

3) Utilizing coherent and non-coherent orthogonal OSTBC: OSTBC is
a well-known MIMO technique with a straightforward encoding design
and low decoding complexity, which can provide full spatial diversity.
However, the application of OSTBC to AmBC has been noted previously
as an open problem \cite{optimal_antenna_Chen_2020}. By leveraging
our linear AmBC MIMO model this can be overcome. Furthermore, noting
that CSI is not always available in AmBC systems, we consider both
coherent and non-coherent OSTBC. When CSI is available, we utilize
coherent OSTBC to enhance the system performance with low complexity
and the general performance metrics for OSTBC under AmBC system have
been given. When CSI is not available, we utilize differential OSTBC
and neither CSIR nor CSIT is needed, which reduces the complexity
and power consumption and outperforms other work using multiple antennas
on the Tag. To our best knowledge, this is the first work using multiple
antennas on the Tag and on Reader without requiring any knowledge
of CSI. For both cases, analysis and simulation results are provided
to verify the performance of our approach.

\subsection{Organization and Notation}

In Section II, an accurate channel model of the AmBC system with a
multiple-antenna Tag and a multiple-antenna Reader is presented. In
Section III, we propose the use of linearization and normalization
to provide a linearized and normalized MIMO channel model for the
proposed AmBC system. Using the proposed AmBC MIMO channel model we
apply coherent and non-coherent OSTBC to AmBC systems and provide
their corresponding detectors in Sections IV and V, respectively.
Section VI provides simulation results and Section VII concludes the
paper.

Bold lower and upper case letters denote vectors and matrices, respectively.
A symbol not in bold font represents a scalar. $\mathbb{E}\left[\cdot\right]$
refers to expectation. $x^{*}$, $\mathfrak{R}\left\{ x\right\} $,
$\mathfrak{I}\left\{ x\right\} $ and $\left|x\right|$ refer to the
conjugate, real part, imaginary part and modulus of a complex scalar
$x$, respectively. $\left\Vert \mathbf{x}\right\Vert $ refers to
the $l_{2}$-norm of a vector $\mathbf{x}$. $\mathbf{X}^{T}$ and
$\mathbf{X}^{H}$ refer to the transpose and conjugate transpose of
a matrix $\mathbf{X}$, respectively. $\mathcal{N}\left(\mu,\sigma^{2}\right)$
and $\mathcal{CN}\left(\mu,\sigma^{2}\right)$ refer to the Gaussian
distribution and the circularly symmetric complex Gaussian distribution
with mean $\mu$ and variance $\sigma^{2}$, respectively. $\stackrel{d}{\rightarrow}$
refers to the convergence in distribution. $Q\left(\cdot\right)$
refers to the Q function.

\section{Ambient Backscatter Communication System}

\subsection{System Model}

\begin{figure}[t]
\centering{}\includegraphics[scale=0.58]{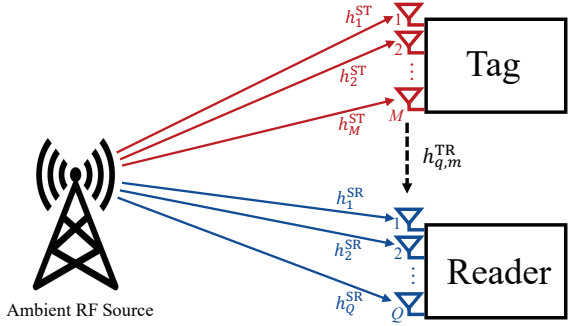}\caption{\label{fig:System-model}System model for an AmBC system with a multiple-antenna
Tag and a multiple-antenna Reader.}
\end{figure}

Consider an AmBC system consisting of an ambient RF source, a passive
Tag equipped with $M$ antennas and a Reader equipped with $Q$ antennas
as shown in Fig. \ref{fig:System-model}. The ambient RF signal is
denoted by $s\left(n\right)$ with symbol period $T_{a}$ where $n=1,2\cdots$
is the symbol index of the ambient source. $s\left(n\right)$ is assumed
random (and may come from different RF sources) and is assumed to
follow a circularly symmetric complex Gaussian distribution $\mathcal{CN}\left(0,P_{s}\right)$,
where $P_{s}$ denotes the average power of the ambient RF signal
\cite{Nocoherent_Qian_2017}. For the $n$th symbol period of the
ambient source, the signal received by the $q$th Reader antenna from
the ambient RF source is expressed as 
\begin{equation}
r_{q}\left(n\right)=A_{\mathrm{SR}}h_{q}^{\mathrm{SR}}s\left(n\right),
\end{equation}
where $A_{\mathrm{SR}}$ and $h_{q}^{\mathrm{SR}}$ respectively denote
the large- and small-scale channel fading between the ambient RF source
and the $q$th Reader antenna, i.e. $h_{q}^{\mathrm{SR}}\sim\mathcal{CN}\left(0,1\right)$.
The signal received by the $m$th Tag antenna from the ambient RF
source is expressed as 
\begin{equation}
t_{m}\left(n\right)=A_{\mathrm{ST}}h_{m}^{\mathrm{ST}}s\left(n\right),
\end{equation}
where $A_{\mathrm{ST}}$ and $h_{m}^{\mathrm{ST}}$ denote the large-
and small-scale channel fading between the ambient RF source and the
$m$th Tag antenna, i.e. $h_{m}^{\mathrm{ST}}\sim\mathcal{CN}\left(0,1\right)$.
The signal backscattered by the $m$th Tag antenna can be expressed
as 
\begin{equation}
d_{m}\left(n\right)=\alpha x_{m}t_{m}\left(n\right),
\end{equation}
where $\alpha$ denotes the hardware implementation loss of the Tag
and $x_{m}$ refers to the symbol transmitted by the $m$th Tag antenna
using backscattering \cite{Signal_detection_multiple_antenna_Tag_Kang_2017}-\nocite{Transceiver_Chen_2020}\cite{optimal_antenna_Chen_2020}.
The total signal received by the $q$th Reader antenna can be expressed
as 
\begin{align}
 & z_{q}\left(n\right)=r_{q}\left(n\right)+\sum_{m=1}^{M}A_{\mathrm{TR}}h_{q,m}^{\mathrm{TR}}d_{m}\left(n\right)+w_{q}\left(n\right)\nonumber \\
= & \left(A_{\mathrm{SR}}h_{q}^{\mathrm{SR}}+\alpha A_{\mathrm{TR}}A_{\mathrm{ST}}\sum_{m=1}^{M}h_{q,m}^{\mathrm{TR}}h_{m}^{\mathrm{ST}}x_{m}\right)s\left(n\right)+w_{q}\left(n\right),\label{eq:zq(n)}
\end{align}
where $A_{\mathrm{TR}}$ and $h_{q,m}^{\mathrm{TR}}$ denote the large-
and small-scale channel fading between the $m$th Tag antenna and
the $q$th Reader antenna, and $w_{q}\left(n\right)$ is the additive
white Gaussian noise (AWGN) at the $q$th Reader antenna, i.e. $h_{q,m}^{\mathrm{TR}}\sim\mathcal{CN}\left(0,1\right)$,
$w_{q}\left(n\right)\sim\mathcal{CN}\left(0,1\right)$.

We assume $h_{q}^{\mathrm{SR}}$, $h_{q,m}^{\mathrm{TR}}$, and $h_{m}^{\mathrm{ST}}$
$\forall$ $m$ and $q$ are independent and identically distributed
(i.i.d.), quasi-static, and frequency flat. We also assume $A_{\mathrm{SR}}\approx A_{\mathrm{ST}}$
since the Tag and the Reader are close to each other. Therefore, we
can normalize \eqref{eq:zq(n)} by $A_{\mathrm{SR}}$ to equivalently
rewrite \eqref{eq:zq(n)} as 
\begin{align}
\bar{z}_{q}\left(n\right) & =\left(h_{q}^{\mathrm{SR}}+\mathbf{h}_{q}^{\mathrm{TR}}\mathbf{G}\mathbf{x}\right)s\left(n\right)+\bar{w}_{q}\left(n\right),\label{eq:normalized Zq(n)}
\end{align}
where we define $\bar{z}_{q}\left(n\right)=z{}_{q}\left(n\right)/A_{\mathrm{SR}}$,
$\bar{w}_{q}\left(n\right)=w_{q}\left(n\right)/A_{\mathrm{SR}}$,
$\mathbf{x}=\left[x_{1},x_{2},...,x_{M}\right]^{T}$, $\mathbf{h}_{q}^{\mathrm{TR}}=\left[h_{q,1}^{\mathrm{TR}},h_{q,2}^{\mathrm{TR}},\ldots,h_{q,M}^{\mathrm{TR}}\right]$,
and $\mathbf{G}=\alpha A_{\mathrm{TR}}\mathrm{diag}\left(h_{1}^{\mathrm{ST}},h_{2}^{\mathrm{ST}},\ldots,h_{M}^{\mathrm{ST}}\right)$.
We assume $\bar{w}_{q}\left(n\right)\sim\mathcal{CN}\left(0,\sigma^{2}\right)$
where $\sigma^{2}$ denotes the normalized noise power and $\mathbf{x}$
is referred to as the Tag transmit signal. It is important to note
that $\left|x_{m}\right|\leq1$ due to the passive backscattered reflection
so that we have 
\begin{equation}
\mathrm{\mathbb{E}}\left[\left\Vert \mathbf{x}\right\Vert ^{2}\right]\leq P_{T}=M,\label{eq:Pt =00003D M}
\end{equation}
where $P_{T}$ denotes the maximum transmit power backscattered from
the Tag. Different from conventional MIMO communications, where $P_{T}$
is a constant, in AmBC $P_{T}$ linearly increases with the number
of Tag antennas $M$. Therefore, using multiple-antenna Tag brings
extra power gain in addition to diversity and multiplexing gain in
conventional MIMO.

Leveraging \eqref{eq:normalized Zq(n)}, we can define direct link
signal-to-noise ratio (SNR) $\gamma_{d}$ as 
\begin{equation}
\gamma_{d}=\frac{P_{s}}{\sigma^{2}},
\end{equation}
which is the ratio of ambient RF signal power and normalized noise
power at the Reader, and define a relative SNR $\Delta\gamma$ as
\begin{equation}
\Delta\gamma=\frac{1}{\alpha^{2}A_{\mathrm{TR}}^{2}},
\end{equation}
which is the ratio of the signal power from the direct link and the
backscatter link.

Since the ambient RF signal $s\left(n\right)$ is unknown, it is challenging
to detect the transmit signal $\mathbf{x}$ from $\bar{z}_{q}\left(n\right)$.
To overcome this challenge, we consider leveraging the statistics
of $\bar{z}_{q}\left(n\right)$ to detect $\mathbf{x}$, as $s\left(n\right)$
is assumed as Gaussian. Because $s\left(n\right)$ and $\bar{w}_{q}\left(n\right)$
are independent circularly-symmetric complex Gaussian random variables,
we can find the distribution of $\bar{z}_{q}\left(n\right)$ as
\begin{align}
\bar{z}_{q}\left(n\right) & \sim\mathcal{CN}\left(0,P_{s}\left|h_{q}^{\mathrm{SR}}+\mathbf{h}_{q}^{\mathrm{TR}}\mathbf{G}\mathbf{x}\right|^{2}+\sigma^{2}\right),
\end{align}
so that we have the first order statistic $\mathrm{\mathbb{E}}\left[\bar{z}_{q}\left(n\right)\right]=0$
and the second order statistic 
\begin{align}
\mathrm{\mathbb{E}}\left[\left|\bar{z}_{q}\left(n\right)\right|^{2}\right] & =P_{s}\left|h_{q}^{\mathrm{SR}}+\mathbf{h}_{q}^{\mathrm{TR}}\mathbf{G}\mathbf{x}\right|^{2}+\sigma^{2}.
\end{align}
In other words, $\mathrm{\mathbb{E}}\left[\bar{z}_{q}\left(n\right)\right]$
contains no information about the transmit signal $\mathbf{x}$, but
$\mathrm{\mathbb{E}}\left[\left|\bar{z}_{q}\left(n\right)\right|^{2}\right]$
has. Therefore, it implies that we should leverage $\mathrm{\mathbb{E}}\left[\left|\bar{z}_{q}\left(n\right)\right|^{2}\right]$
to detect the transmit signal $\mathbf{x}$.

\subsection{Averaging Process}

To estimate $\mathbb{E}\left[\left|\bar{z}_{q}\left(n\right)\right|^{2}\right]$
for detecting the transmit signal $\mathbf{x}$, we consider $N$
symbols of $\bar{z}_{q}\left(n\right)$, i.e. $\bar{z}_{q}\left(1\right)$,
$\bar{z}_{q}\left(2\right)$, ..., and $\bar{z}_{q}\left(N\right)$,
and take them as $N$ i.i.d. samples. In other words, for each transmit
signal $\mathbf{x}$, we use $N$ symbols of $\bar{z}_{q}\left(n\right)$
to detect it. Therefore, the Tag transmits signal $\mathbf{x}$ with
a symbol period $T=NT_{a}$, so that the symbol rate of the signal
transmitted by the Tag is $N$ times smaller than the ambient signal.
Following previous methods \cite{thin_Liu_2013}, we average the power
of $\bar{z}_{q}\left(n\right)$ as 
\begin{equation}
\bar{y}_{q}=\frac{1}{N}\sum_{n=1}^{N}\mathop{\left|\bar{z}_{q}\left(n\right)\right|^{2}},\label{eq:averaging process}
\end{equation}
so as to estimate $\mathbb{E}\left[\left|\bar{z}_{q}\left(n\right)\right|^{2}\right]$.
The resulting signal $\bar{y}_{q}$ is distributed as 
\begin{equation}
\bar{y}_{q}\sim\frac{P_{s}\left|h_{q}^{\mathrm{SR}}+\mathbf{h}_{q}^{\mathrm{TR}}\mathbf{G}\mathbf{x}\right|^{2}+\sigma^{2}}{2N}\chi^{2}\left(2N\right),\label{accurate}
\end{equation}
where $\chi^{2}\left(2N\right)$ refers to the chi-square distribution
with $2N$ degrees of freedom. As $N\rightarrow\infty$, we have $\chi^{2}\left(2N\right)\stackrel{d}{\rightarrow}\mathcal{N}\left(2N,4N\right)$
(normally $N=30$ is adequate for most applications \cite{Nocoherent_Qian_2017},
\cite{Probability_Papoulis_2002}), and accordingly we have 
\begin{equation}
\bar{y}_{q}\sim\left(P_{s}\left|h_{q}^{\mathrm{SR}}+\mathbf{h}_{q}^{\mathrm{TR}}\mathbf{G}\mathbf{x}\right|^{2}+\sigma^{2}\right)\mathcal{N}\left(1,\frac{1}{N}\right).\label{yq}
\end{equation}
Since $\bar{y}_{q}$ has mean $P_{s}\left|h_{q}^{\mathrm{SR}}+\mathbf{h}_{q}^{\mathrm{TR}}\mathbf{G}\mathbf{x}\right|^{2}+\sigma^{2}$,
it can be decomposed into three parts 
\begin{equation}
\bar{y}_{q}=f_{q}\left(\mathbf{x}\right)+c_{q}+\bar{n}_{q},\label{eq:channel model}
\end{equation}
where $f_{q}\left(\mathbf{x}\right)$ refers to the signal and is
given by 
\begin{align}
f_{q}\left(\mathbf{x}\right) & =P_{s}\left|h_{q}^{\mathrm{SR}}+\mathbf{h}_{q}^{\mathrm{TR}}\mathbf{G}\mathbf{x}\right|^{2}+\sigma^{2}-\left(P_{s}\left|h_{q}^{\mathrm{SR}}\right|^{2}+\sigma^{2}\right)\nonumber \\
 & =2P_{s}\mathfrak{R}\left\{ h_{q}^{\mathrm{SR*}}\mathbf{h}_{q}^{\mathrm{TR}}\mathbf{G}\mathbf{x}\right\} +P_{s}\left|\mathbf{h}_{q}^{\mathrm{TR}}\mathbf{G}\mathbf{x}\right|^{2}.\label{f(x)}
\end{align}
$c_{q}$ refers to a bias containing no information about the transmit
signal and is a constant given by 
\begin{equation}
c_{q}=P_{s}\left|h_{q}^{\mathrm{SR}}\right|^{2}+\sigma^{2},\label{eq:bias}
\end{equation}
where $c_{q}+f_{q}\left(\mathbf{x}\right)=P_{s}\left|h_{q}^{\mathrm{SR}}+\mathbf{h}_{q}^{\mathrm{TR}}\mathbf{G}\mathbf{x}\right|^{2}+\sigma^{2}$.
$\bar{n}_{q}$ refers to the noise given by 
\begin{equation}
\bar{n}_{q}\sim\left(P_{s}\left|h_{q}^{\mathrm{SR}}+\mathbf{h}_{q}^{\mathrm{TR}}\mathbf{G}\mathbf{x}\right|^{2}+\sigma^{2}\right)\mathcal{N}\left(0,\frac{1}{N}\right).\label{nq}
\end{equation}

It is also worth noting that the received signals must be symbol-synchronized
in order to determine when the Tag starts to backscatter signal. In
\cite{AmBC_FM_Daskalakis_2017}, a preamble sequence is added to the
head of the bit stream. At the receiver, cross correlation of the
known preamble symbol sequence and received signal is calculated.
The starting point of the bit stream is defined as the point that
maximizes the cross correlation.

\subsection{Challenges to Use AmBC Channel Model}

Focusing on the previous channel model for AmBC, \eqref{eq:channel model},
we find that there are two challenges to use the AmBC channel model.

\subsubsection{Nonlinear Channel}

The signal $f_{q}\left(\mathbf{x}\right)$ is a quadratic function
of $\mathbf{x}$, which is different from the linear function of $\mathbf{x}$
in the conventional linear MIMO channel model. In addition, there
is a constant bias $c_{q}$ having no information of the transmit
signal $\mathbf{x}$, which arises from the averaging process.

\subsubsection{Dependent Noise}

The noise $\bar{n}_{q}$ has a zero mean and a variance, denoted as
$\varsigma_{q}^{2}$, which is given by 
\begin{equation}
\varsigma_{q}=\frac{P_{s}\left|h_{q}^{\mathrm{SR}}+\mathbf{h}_{q}^{\mathrm{TR}}\mathbf{G}\mathbf{x}\right|^{2}+\sigma^{2}}{\sqrt{N}}.\label{eq:non-independent noise}
\end{equation}
We find that the variance of $\bar{n}_{q}$ depends on the transmit
signal ${\bf x}$, and the channels $h_{q}^{\mathrm{SR}}$, $\mathbf{h}_{q}^{\mathrm{TR}}$,
and $\mathbf{G}$, which is significantly different from the conventional
AWGN.

The two challenges make the detection of transmit signal $\mathbf{x}$
difficult and increase the detection complexity, especially for OSTBC
detection as shown in Sections IV and V.

\section{AmBC Channel Linearization}

In this section, we propose two techniques, namely AmBC channel linearization
and noise normalization, to overcome the two challenges of using the
AmBC channel model \eqref{eq:channel model}, so that conventional
approaches in MIMO communications such as OSTBC can be applied to
AmBC.

\subsection{Linearization}

The direct link (ambient RF source to Reader) has a significantly
higher channel gain than the indirect link (ambient RF source to Tag
and then backscattered to Reader). This corresponds to the fact that
the relative SNR $\Delta\gamma$ in practical AmBC systems is usually
higher than $30$ dB, which is confirmed by simulation in Section
VI. Therefore, we have that 
\begin{equation}
\left|\mathbf{h}_{q}^{\mathrm{TR}}\mathbf{G}\mathbf{x}\right|\ll\left|h_{q}^{\mathrm{SR}}\right|.\label{eq:AP2}
\end{equation}
We rewrite $\left|h_{q}^{\mathrm{SR}}+\mathbf{h}_{q}^{\mathrm{TR}}\mathbf{G}\mathbf{x}\right|^{2}$
as 
\begin{align}
 & \left|h_{q}^{\mathrm{SR}}+\mathbf{h}_{q}^{\mathrm{TR}}\mathbf{G}\mathbf{x}\right|^{2}\nonumber \\
 & =\left|h_{q}^{\mathrm{SR}}\right|^{2}\left(\left(1+\mathfrak{R}\left\{ \frac{\mathbf{h}_{q}^{\mathrm{TR}}\mathbf{G}\mathbf{x}}{h_{q}^{\mathrm{SR}}}\right\} \right)^{2}+\mathfrak{I}\left\{ \frac{\mathbf{h}_{q}^{\mathrm{TR}}\mathbf{G}\mathbf{x}}{h_{q}^{\mathrm{SR}}}\right\} ^{2}\right).\label{eq:Taylor1}
\end{align}
From \eqref{eq:AP2}, we have that $\mathfrak{R}\left\{ \frac{\mathbf{h}_{q}^{\mathrm{TR}}\mathbf{G}\mathbf{x}}{h_{q}^{\mathrm{SR}}}\right\} \rightarrow0$
and $\mathfrak{I}\left\{ \frac{\mathbf{h}_{q}^{\mathrm{TR}}\mathbf{G}\mathbf{x}}{h_{q}^{\mathrm{SR}}}\right\} \rightarrow0$.
Subsequently, by taking the Taylor expansion of $\left(1+\mathfrak{R}\left\{ \frac{\mathbf{h}_{q}^{\mathrm{TR}}\mathbf{G}\mathbf{x}}{h_{q}^{\mathrm{SR}}}\right\} \right)^{2}+\mathfrak{I}\left\{ \frac{\mathbf{h}_{q}^{\mathrm{TR}}\mathbf{G}\mathbf{x}}{h_{q}^{\mathrm{SR}}}\right\} ^{2}$,
we can ignore the second order terms so as to approximate \eqref{eq:Taylor1}
as 
\begin{equation}
\left|h_{q}^{\mathrm{SR}}+\mathbf{h}_{q}^{\mathrm{TR}}\mathbf{G}\mathbf{x}\right|^{2}\thickapprox\left|h_{q}^{\mathrm{SR}}\right|^{2}+2\mathfrak{R}\left\{ h_{q}^{\mathrm{SR*}}\mathbf{h}_{q}^{\mathrm{TR}}\mathbf{G}\mathbf{x}\right\} .\label{eq:AP1}
\end{equation}

Therefore, according to the definition of $f_{q}\left(\mathbf{x}\right)$
\eqref{f(x)}, we can approximate $f_{q}\left(\mathbf{x}\right)$
as 
\begin{equation}
f_{q}\left(\mathbf{x}\right)\approx2P_{s}\mathfrak{R}\left\{ h_{q}^{\mathrm{SR*}}\mathbf{h}_{q}^{\mathrm{TR}}\mathbf{G}\mathbf{x}\right\} .\label{approximate f(x)}
\end{equation}

To quantify the error for omitting $P_{s}\left|\mathbf{h}_{q}^{\mathrm{TR}}\mathbf{G}\mathbf{x}\right|^{2}$
in the information signal \eqref{f(x)}, we express its error relative
to the total information signal as
\begin{equation}
\epsilon=\frac{\mathbb{E}\left[P_{s}\left|\mathbf{h}_{q}^{\mathrm{TR}}\mathbf{G}\mathbf{x}\right|^{2}\right]}{\mathbb{E}\left[\left|2P_{s}\mathfrak{R}\left\{ h_{q}^{\mathrm{SR*}}\mathbf{h}_{q}^{\mathrm{TR}}\mathbf{G}\mathbf{x}\right\} +P_{s}\left|\mathbf{h}_{q}^{\mathrm{TR}}\mathbf{G}\mathbf{x}\right|^{2}\right|\right]}.\label{eq: approximation error}
\end{equation}
 It is shown in Section VI that this error is negligible.

The constant bias term $c_{q}$ in the channel model \eqref{eq:channel model}
can also be removed. According to \eqref{eq:channel model}, one way
to estimate $c_{q}$ is to have the Tag not transmit anything in the
first symbol period (equivalently the transmit signal is $\mathbf{x}=\mathbf{0}$).
Since $f_{q}(\mathbf{0})=0$, at the Reader we have that 
\begin{equation}
\bar{y}_{q,\mathbf{x}=\mathbf{0}}=f_{q}(\mathbf{0})+c_{q}+\bar{n}_{q}=c_{q}+\bar{n}_{q}.\label{eq:estimate_cq}
\end{equation}
Therefore, we can select $\bar{y}_{q,\mathbf{x}=\mathbf{0}}$ as the
estimated bias $c_{q}$ since the noise term $\bar{n}_{q}$ is small.
This estimation approach has also been used previously \cite{AmBC_FM_Daskalakis_2017},
\cite{Nocoherent_Qian_2017}, \cite{Transceiver_Chen_2020}, \cite{Semi_coherent_Qian_2017}. 

It is also important to consider the estimation errors of the bias
$c_{q}$. As shown in equation \eqref{eq:estimate_cq}, the estimation
error is the noise term $\bar{n}_{q}$. Such estimation error will
degrade detection performance in AmBC. However, from equation \eqref{nq},
we can minimize the estimation error by increasing the number of samples
$N$. By using a large value of $N$ to estimate $c_{q}$, the estimation
error can be reduced. In the remainder of the paper we assume perfect
estimation of $c_{q}$ for simplicity. 

In summary, based on the approximation \eqref{approximate f(x)} and
removing bias $c_{q}$, the nonlinear AmBC channel can be entirely
linearized.

\subsection{Noise Normalization}

As shown in \eqref{eq:non-independent noise}, the variance of the
noise depends on the transmit signal ${\bf x}$, and the channels
$h_{q}^{\mathrm{SR}}$, $\mathbf{h}_{q}^{\mathrm{TR}}$, and $\mathbf{G}$.
This makes the detection of ${\bf x}$ difficult since 1) the noise
$\bar{n}_{q}$ is correlated with the transmit signal ${\bf x}$,
and 2) the noise $\bar{n}_{q}$ has different power levels at different
Reader antennas.

To solve this difficulty, we first omit the term $\mathbf{h}_{q}^{\mathrm{TR}}\mathbf{G}\mathbf{x}$
in \eqref{eq:non-independent noise} according to \eqref{eq:AP2}
so that the standard deviation of the noise $\varsigma_{q}$ can be
approximately written as 
\begin{equation}
\varsigma_{q}\approx\frac{P_{s}\left|h_{q}^{\mathrm{SR}}\right|^{2}+\sigma^{2}}{\sqrt{N}}=\frac{c_{q}}{\sqrt{N}}.\label{eq:decorrelated noise}
\end{equation}
This approximation allows us to the treat the noise as being uncorrelated
to the transmit signal $\mathbf{x}$. In addition we have tested,
numerically, that omitting the term $\mathbf{h}_{q}^{\mathrm{TR}}\mathbf{G}\mathbf{x}$
in \eqref{eq:non-independent noise} has nearly no impact on detection.

With \eqref{approximate f(x)}, \eqref{eq:decorrelated noise}, and
removing bias $c_{q}$, we can rewrite our AmBC channel model \eqref{eq:channel model}
as 
\begin{equation}
\bar{y}_{q}-c_{q}=2P_{s}\mathfrak{R}\left\{ h_{q}^{\mathrm{SR*}}\mathbf{h}_{q}^{\mathrm{TR}}\mathbf{G}\mathbf{x}\right\} +\frac{c_{q}}{\sqrt{N}}n_{q}.\label{eq:AP channel 1}
\end{equation}
where $n_{q}\sim\mathcal{N}\left(0,1\right)$. However, the approximate
noise \eqref{eq:decorrelated noise} still has power levels dependent
on $\left|h_{q}^{\mathrm{SR}}\right|^{2}$ at different Reader antennas,
which is still different from the conventional i.i.d. AWGN. Therefore,
we normalize \eqref{eq:AP channel 1} by $c_{q}/\sqrt{N}$ by defining
\begin{equation}
y_{q}=\sqrt{N}\frac{\bar{y}_{q}-c_{q}}{c_{q}},\label{eq:normalization}
\end{equation}
so that we have 
\begin{equation}
y_{q}=\frac{2\sqrt{N}\gamma_{d}}{\gamma_{d}\left|h_{q}^{\mathrm{SR}}\right|^{2}+1}\mathfrak{R}\left\{ h_{q}^{\mathrm{SR*}}\mathbf{h}_{q}^{\mathrm{TR}}\mathbf{G}\mathbf{x}\right\} +n_{q}.\label{eq:AmBC linear Model 1}
\end{equation}
which is now a linear channel model with i.i.d. AWGN.

In summary, by approximating $f_{q}\left(\mathbf{x}\right)$ \eqref{approximate f(x)},
removing bias \eqref{eq:estimate_cq}, approximating the noise \eqref{eq:decorrelated noise},
and normalizing the noise power \eqref{eq:normalization}, we can
approximate the original channel model in AmBC \eqref{eq:channel model}
as a linear channel model with i.i.d. AWGN \eqref{eq:AmBC linear Model 1}.
This channel model approximation is accurate for most practical configurations
in AmBC as shown later in Section VI.

\subsection{BPSK Modulation}

In this paper, we focus on using BPSK modulation for the transmit
signal $\mathbf{x}$, i.e. $x_{m}=\pm1$. For BPSK modulation (or
other real constellations), we can rewrite \eqref{eq:AmBC linear Model 1}
in a compact form as 
\begin{equation}
y_{q}=\mathbf{h}_{q}\mathbf{x}+n_{q},\label{eq:linearized and normalized MISO h model}
\end{equation}
where $\mathbf{h}_{q}=\left[h_{q,1},h_{q,2},\ldots,h_{q,M}\right]$
and $h_{q,m}$ is given by 
\begin{equation}
h_{q,m}=2\alpha A_{\mathrm{TR}}\sqrt{N}\frac{\gamma_{d}\mathfrak{R}\left\{ h_{q}^{\mathrm{SR*}}h_{q,m}^{\mathrm{TR}}h_{m}^{\mathrm{ST}}\right\} }{\gamma_{d}\left|h_{q}^{\mathrm{SR}}\right|^{2}+1}.\label{eq:hq}
\end{equation}

For the proposed AmBC system with multiple-antenna Tag, the receive
SNR at the Reader is given by 
\begin{align}
\frac{\mathbb{E}\left[\left|\mathbf{h}_{q}\mathbf{x}\right|^{2}\right]}{\mathbb{E}\left[\left|n_{q}\right|^{2}\right]} & =\mathbb{E}\left[\left\Vert \mathbf{h}_{q}\right\Vert {}^{2}\right]\nonumber \\
 & =\frac{4MN}{\Delta\gamma}\mathbb{E}\left[\left(\frac{\gamma_{d}\mathfrak{R}\left\{ h_{1}^{\mathrm{SR*}}h_{1,1}^{\mathrm{TR}}h_{1}^{\mathrm{ST}}\right\} }{\gamma_{d}\left|h_{1}^{\mathrm{SR}}\right|^{2}+1}\right)^{2}\right],\label{eq:new SNR}
\end{align}
which for all $q$ is identical. It is important to note that using
multiple antennas at the Tag can increase the receive SNR at the Reader
by $M$ times, which arises from \eqref{eq:Pt =00003D M}. To highlight
this benefit of the multiple-antenna Tag over a single-antenna Tag,
we define as a reference the receive SNR for a single-antenna Tag
AmBC, which is denoted as $\gamma_{R}$ and given by 
\begin{equation}
\gamma_{R}=\frac{4N}{\Delta\gamma}\mathbb{E}\left[\left(\frac{\gamma_{d}\mathfrak{R}\left\{ h_{1}^{\mathrm{SR*}}h_{1,1}^{\mathrm{TR}}h_{1}^{\mathrm{ST}}\right\} }{\gamma_{d}\left|h_{1}^{\mathrm{SR}}\right|^{2}+1}\right)^{2}\right].\label{eq:yr-siso}
\end{equation}
Accordingly, in the simulation results in Section VI, the BER performance
is evaluated versus $\gamma_{R}$. When $\gamma_{d}\rightarrow0$,
we have 
\begin{equation}
\gamma_{R}=\frac{4N\gamma_{d}^{2}}{\Delta\gamma}\mathbb{E}\left[\mathfrak{R}^{2}\left\{ h_{1}^{\mathrm{SR*}}h_{1,1}^{\mathrm{TR}}h_{1}^{\mathrm{ST}}\right\} \right],
\end{equation}
which means that $\gamma_{R}$ increases with $\gamma_{d}$ quadratically.
However, when $\gamma_{d}\rightarrow\infty$, we have 
\begin{equation}
\gamma_{R}=\frac{4N}{\Delta\gamma}\mathbb{E}\left[\mathfrak{R}^{2}\left\{ \frac{h_{1,1}^{\mathrm{TR}}h_{1}^{\mathrm{ST}}}{h_{1}^{\mathrm{SR}}}\right\} \right],\label{eq:yr limit}
\end{equation}
which means that $\gamma_{R}$ cannot go to infinity with increasing
$\gamma_{d}$ and agrees with the expected result in practice.

We collect $y_{q}$ for all $q$ into $\mathbf{y}=\left[y_{1},y_{2},\ldots,y_{Q}\right]^{T}$,
$n_{q}$ for all $q$ into $\mathbf{n}=\left[n_{1},n_{2},\ldots,n_{Q}\right]^{T}$,
and $\mathbf{h}_{q}$ for all $q$ into $\mathbf{H}=\left[\mathbf{h}_{1}^{T},\mathbf{h}_{2}^{T},\ldots,\mathbf{h}_{Q}^{T}\right]^{T}$,
so that we can construct a linearized and normalized MIMO channel
model for AmBC as 
\begin{equation}
\mathbf{y}=\mathbf{H}\mathbf{x}+\mathbf{n}.\label{eq:Full MIMO}
\end{equation}
Based on the MIMO channel model \eqref{eq:Full MIMO}, we can apply
conventional MIMO communication approaches to AmBC to enhance its
performance. In the following sections, we apply coherent and non-coherent
OSTBC to enhance the backscattered signal detection performance in
AmBC.

\section{Coherent OSTBC}

We apply OSTBC \cite{OSTBC_1999} in the proposed AmBC system to enhance
performance because it has a straightforward design, low decoding
complexity and low power consumption. Since ambient signals are unknown,
it has been an open problem to utilize OSTBC in AmBC systems \cite{optimal_antenna_Chen_2020}.
By leveraging our linearized and normalized MIMO channel model \eqref{eq:Full MIMO},
OSTBC can be straightforwardly applied to AmBC systems. Using OSTBC
requires CSIR, but does not require CSIT so that there is no need
for any feedback. The basic principle of OSTBC is that $M$ symbols
are coded across $M$ antennas and also over $M$ symbol periods,
so that a multiplexing gain of unity is maintained. It should be noted
that real orthogonal design exists if and only if $M=2,4,8$.

\subsection{Encoding Algorithm}

Consider $M$ bits $\left(M=2,4,8\right)$, $b_{Mi-M+1},...,b_{Mi}$,
that are to be transmitted as the $i$th block. Using BPSK, $b_{Mi-M+1},...,b_{Mi}$
are modulated to $M$ symbols $u_{Mi-M+1},...,u_{Mi}\in\left\{ \pm1\right\} $.
According to \cite{OSTBC_1999}, the $M$ symbols are encoded into
a block $\mathbf{X}_{i}=\left[\mathbf{x}_{Mi-M+1},\mathbf{x}_{Mi-M+2},...,\mathbf{x}_{Mi}\right]$,
where $\mathbf{x}_{Mi}$ denotes the signals transmitted by $M$ Tag
antennas at symbol period $Mi$ and satisfies 
\begin{equation}
\mathbf{X}_{i}^{T}\mathbf{X}_{i}=\left(u_{Mi-M+1}^{2}+...+u_{Mi}^{2}\right)\mathbf{I}_{M},\label{eq:encoding_OSTBC}
\end{equation}
where $\mathbf{I}_{M}$ represents the $M\times M$ identity matrix.

It terms of implementing this at the AmBC Tag, it should be noted
that by tuning the load impedance of the Tag, the transmitted OSTBC
symbol can be mapped to the reflection coefficient of the load impedance
\cite{survery_VanHuynh_2018}. In addition, a low-power micro-controller
on the Tag can save and process the data needed to form a block \cite{MIMO_RFID_Boyer_2014}.

\subsection{Coherent Detection}

Since encoding and detecting $\mathbf{X}_{i}$ do not rely on the
other blocks, we simplify the analysis by only considering the first
block $\mathbf{X}_{1}=\left[\mathbf{x}_{1},\mathbf{x}_{2},...,\mathbf{x}_{M}\right]$.
Two coherent detectors are proposed to detect the first signal vector
$\mathbf{u}_{1}=[u_{1},u_{2},...,u_{M}]^{T}$. The first detector
is the optimal ML detector based on the accurate AmBC channel model
\eqref{eq:channel model}. The second detector is a linear detector
with low complexity based on the linearized and normalized MIMO channel
model \eqref{eq:Full MIMO}. The two detectors are compared with each
other in the numerical experiments as shown in Section VI to show
the accuracy of proposed MIMO channel model.

Recall that for each transmit signal $\mathbf{x}$, we use $N$ symbols
of $\bar{z}_{q}\left(n\right)$ to detect it through the averaging
process as shown in Section II.B. Therefore, it takes $MN$ symbols
of $\bar{z}_{q}\left(n\right)$ to decode the block $\mathbf{X}_{1}$.
For example, as per \eqref{eq:averaging process}, $\bar{z}_{q}\left(1\right)$,
..., $\bar{z}_{q}\left(N\right)$ are used to achieve $\bar{y}_{q,1}$
and $\bar{z}_{q}\left(\left(M-1\right)N+1\right)$, ..., $\bar{z}_{q}\left(MN\right)$
are used to achieve $\bar{y}_{q,M}$. To explicitly show the dependence
on the symbol periods, we add extra subscripts for different terms
such as signals and noise. For example, $\bar{y}_{q,1}$ and $\bar{y}_{q,M}$
denote the signal received by the $q$th Reader antenna at symbol
periods 1 and $M$, respectively.

\subsubsection{Optimal ML Detector}

Based on the accurate AmBC channel model \eqref{eq:channel model},
we apply the optimal ML detector to detect $\mathbf{u}_{1}$. In ML
detector, we maximize the posteriori probability density function
(PDF), $\mathit{f}\left(\bar{\mathbf{Y}}_{1}\mid\mathbf{X}_{1}\right)$,
where $\bar{\mathbf{Y}}_{1}$ is $Q\times M$ matrix with the $\left(q,j\right)$th
entry being $\bar{y}_{q,j}$ for $j=1,2,...,M$. Since $\bar{y}_{q,j}$
at different Reader antennas and different symbol periods are independent
given $\mathbf{X}_{1}$, we can simplify the ML detector as

\begin{equation}
\tilde{\mathbf{u}}_{1}=\underset{u_{1},...,u_{M}\in\left\{ \pm1\right\} }{\textrm{argmax}}\mathbf{\stackrel[\mathit{q=\mathrm{1}}]{\mathit{Q}}{\prod}}\prod_{j=1}^{M}\frac{1}{\varsigma_{q,j}}e^{-\left|\frac{\bar{y}_{q,j}-f_{q}\left(\mathbf{x}_{j}\right)-c_{q}}{\varsigma_{q,j}}\right|^{2}},\label{eq:ML detector}
\end{equation}
where we leverage the Gaussian distribution of the noise \eqref{nq}.
It should be noted that $\varsigma_{q,j}$ relies on $\mathbf{x}_{j}$
as shown \eqref{eq:non-independent noise}. The optimal ML detector
is computationally intricate since it needs to compute the exponential
function and it cannot separately decode $u_{1}$ to $u_{M}$.

\subsubsection{Linear Detector}

To reduce the computational complexity of the optimal ML detector,
we can use a linear detector to decode $u_{1}$ to $u_{M}$ \cite{OSTBC_1999,alamouti_1998}.
However, it is not applicable under the accurate AmBC channel model
\eqref{eq:channel model} because 1) the channel is nonlinear and
2) the noise is dependent on the signals and channels. To solve this
problem, we leverage the linearized and normalized MIMO channel model
\eqref{eq:Full MIMO} and then apply the linear detector.

Specifically, leveraging \eqref{eq:linearized and normalized MISO h model},
we have that 
\begin{equation}
y_{q,j}=\mathbf{h}_{q}\mathbf{x}_{j}+n_{q,j},\:j=1,2,...,M,
\end{equation}
where $y_{q,j}$ can be achieved by linearizing and normalizing $\bar{y}_{q,j}$,
i.e. $y_{q,j}=\sqrt{N}\left(\bar{y}_{q,j}-c_{q}\right)/c_{q}$ and
$\mathbf{h}_{q}$ is given by \eqref{eq:hq}. According to the linear
detector \cite{OSTBC_1999,alamouti_1998}, we first compute $\mathbf{v}_{q}$
by $\mathbf{v}_{q}=\left[v_{q,1},v_{q,2},...,v_{q,M}\right]^{T}=\mathbf{\Lambda}^{T}\left[y_{q,\mathrm{1}},y_{q,\mathrm{2}},...,y_{q,M}\right]^{T}$
where $\mathbf{\Lambda}$ is the matrix that satisfies $\mathbf{\Lambda}\mathbf{u}_{1}=\left[\mathbf{h}_{q}\mathbf{x}_{1},\mathbf{h}_{q}\mathbf{x}_{2},...,\mathbf{h}_{q}\mathbf{x}_{M}\right]^{T}$.
Then we detect $u_{1}$ to $u_{M}$ by 
\begin{equation}
\tilde{u}_{j}=\begin{cases}
1 & ,\textrm{if}\:\stackrel[q=1]{Q}{\sum}v_{q,j}\geqslant0\\
-1 & ,\textrm{if}\:\stackrel[q=1]{Q}{\sum}v_{q,j}<0
\end{cases},\:j=1,2,...,M.\label{eq:linear detector}
\end{equation}
It should be noted that the decision statistic in \eqref{eq:linear detector}
is simplified since BPSK modulation is used. A more general decision
statistic for other modulations can be found in \cite{OSTBC_1999}.
\begin{figure}[t]
\begin{centering}
\subfloat[Coherent OSTBC using optimal detector]{\begin{centering}
\includegraphics[scale=0.325]{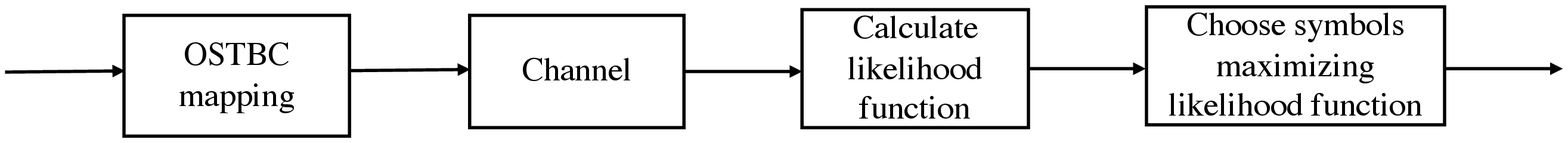}
\par\end{centering}
\centering{}}
\par\end{centering}
\centering{}\subfloat[Coherent OSTBC using linear detector]{\centering{}\includegraphics[scale=0.325]{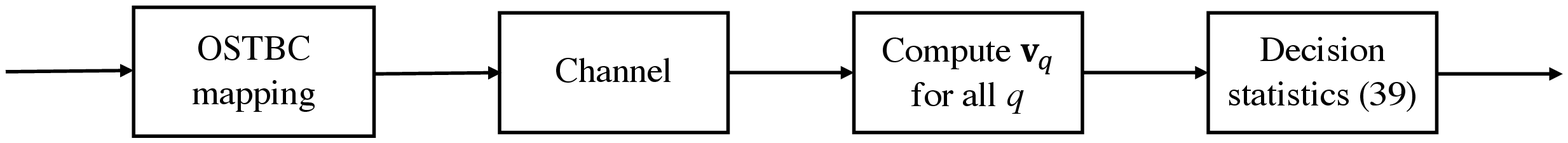}}\caption{\label{fig:Block_diagrams_coherent_STBC}Block diagrams for our proposed
coherent OSTBC systems.}
\end{figure}

To conclude, leveraging the proposed linearized and normalized MIMO
channel model, we can apply linear detection to separately detect
$u_{1}$ to $u_{M}$ and avoid computing the exponential function.
Therefore the computational complexity for detection is greatly reduced.
The block diagrams for coherent OSTBC using optimal and linear detectors
are shown in Fig. \ref{fig:Block_diagrams_coherent_STBC}.

\subsubsection{Equivalence}

It is worthwhile to show the approximate equivalence between the optimal
ML detector based on \eqref{eq:channel model} and the linear detector
based on \eqref{eq:Full MIMO}.

Using the approximated noise \eqref{eq:decorrelated noise}, we can
simplify \eqref{eq:ML detector} as 
\begin{equation}
\tilde{\mathbf{u}}_{1}=\underset{u_{1},...,u_{M}\in\left\{ \pm1\right\} }{\textrm{argmin}}\mathbf{\stackrel[\mathit{q=\mathrm{1}}]{\mathit{Q}}{\sum}}\sum_{j=1}^{M}\left|\sqrt{N}\frac{\bar{y}_{q,j}-f_{q}\left(\mathbf{x}_{j}\right)-c_{q}}{c_{q}}\right|^{2}.\label{eq:ML detector based ap noise}
\end{equation}
Using the channel linearization and the noise normalization, we can
further simplify \eqref{eq:ML detector based ap noise} as 
\begin{equation}
\tilde{\mathbf{u}}_{1}=\underset{u_{1},...,u_{M}\in\left\{ \pm1\right\} }{\textrm{argmin}}\mathbf{\stackrel[\mathit{q=\mathrm{1}}]{\mathit{Q}}{\sum}}\sum_{j=1}^{M}\left|y_{q,j}-\mathbf{h}_{q}\mathbf{x}_{j}\right|^{2},\label{eq:minimum distance detector}
\end{equation}
which is a minimum distance detector \cite{goldsmith_2005}. As shown
previously \cite{OSTBC_1999}, \cite{alamouti_1998}, the minimum
distance detector \eqref{eq:minimum distance detector} is equivalent
to the linear detector \eqref{eq:linear detector}.

It can be concluded that using the linear detector with the linearized
and normalized MIMO channel model \eqref{eq:Full MIMO} is equivalent
to the optimal ML detector with the accurate AmBC channel model \eqref{eq:channel model}.
A numerical comparison for those two detectors is provided in Section
VI to verify the equivalence.

\subsection{BER Analysis}

Leveraging the linearized and normalized MIMO channel \eqref{eq:Full MIMO},
we also provide a BER analysis for the coherent OSTBC with BPSK.

\subsubsection{Single-Antenna Tag Single-Antenna Reader}

We first consider the conventional SISO (i.e. $M=1$ and $Q=1$) AmBC
system using BPSK as a comparison. Using \eqref{eq:linearized and normalized MISO h model},
we have that 
\begin{equation}
y_{1}=2\alpha A_{\mathrm{TR}}\sqrt{N}\gamma_{d}\frac{\mathfrak{R}\left\{ h_{1}^{\mathrm{SR*}}h_{1,1}^{\mathrm{TR}}h_{1}^{\mathrm{ST}}\right\} }{\gamma_{d}\left|h_{1}^{\mathrm{SR}}\right|^{2}+1}x_{1}+n_{1}.
\end{equation}
Given the channel realizations $h_{1}^{\mathrm{SR}}$, $h_{1}^{\mathrm{ST}}$,
and $h_{1,1}^{\mathrm{TR}}$, it is straightforward to derive the
BER for the SISO AmBC as 
\begin{equation}
P_{\mathrm{SISO}}=Q\left(2\alpha A_{\mathrm{TR}}\sqrt{N}\gamma_{d}\left|\frac{\mathfrak{R}\left\{ h_{1}^{\mathrm{SR*}}h_{1,1}^{\mathrm{TR}}h_{1}^{\mathrm{ST}}\right\} }{\gamma_{d}\left|h_{1}^{\mathrm{SR}}\right|^{2}+1}\right|\right).\label{eq:PSISO}
\end{equation}

\subsubsection{Multiple-Antenna Tag Multiple-Antenna Reader}

For the multiple-antenna Tag multiple-antenna Reader AmBC system,
we start by noting that $\stackrel[q=1]{Q}{\sum}v_{q,1}$ given $u_{1}$
is distributed as $\stackrel[q=1]{Q}{\sum}v_{q,1}\sim\mathcal{N}\left(u_{1}\stackrel[q=1]{Q}{\sum}\left\Vert \mathbf{h}_{q}\right\Vert ^{2},\stackrel[q=1]{Q}{\sum}\left\Vert \mathbf{h}_{q}\right\Vert ^{2}\right).$
Given the channel realizations $\mathbf{h}_{q}$, from \eqref{eq:linear detector},
it is straightforward to derive that the BER of AmBC using the linear
detector with BSPK is given by 
\begin{align}
 & P_{\mathrm{Linear}}\nonumber \\
 & =Q\left(2\alpha A_{\mathrm{TR}}\sqrt{N}\gamma_{d}\sqrt{\stackrel[q=1]{Q}{\sum}\sum_{m=1}^{M}\left|\frac{\mathfrak{R}\left\{ h_{q}^{\mathrm{SR*}}h_{q,m}^{\mathrm{TR}}h_{m}^{\mathrm{ST}}\right\} }{\gamma_{d}\left|h_{q}^{\mathrm{SR}}\right|^{2}+1}\right|^{2}}\right).\label{eq:PAlamouti}
\end{align}
Therefore, we can deduce that using multiple-antenna Tag and multiple-antenna
Reader can effectively decrease the BER compared with conventional
SISO AmBC. The BERs of the MIMO AmBC and the SISO AmBC are numerically
compared in Section VI.

\section{Non-Coherent OSTBC}

The limitation of coherent OSTBC is that it requires CSI at the Reader.
Although methods have been demonstrated that provide CSI in AmBC \cite{Blind_channel_estimation_Ma_2018}-\nocite{Backfi_Bharadia_2015,Machine_CSI_mA_2018,Sparse_CSI_kIM_2018,CSI_massive_Reader_Zhao_2019,multitag_channel_estimation_Mishra_2019}\cite{joint_CSI_Darsena_2018},
estimating the channel is not desirable in AmBC. For example a blind
channel estimation method based on expectation maximization (EM)-based
estimator \cite{Blind_channel_estimation_Ma_2018}, \cite{Machine_CSI_mA_2018},
has been proposed to obtain CSI. While it is effective, the channel
estimation process inevitably increases the complexity of AmBC system
design as well as power consumption and it is therefore useful to
consider methods that do not need CSI \cite{Nocoherent_Qian_2017}. 

We propose using non-coherent differential OSTBC to remove the need
for CSI. Differential OSTBC is a transmission scheme for exploiting
diversity given by multiple antennas when neither the transmitter
nor the receiver requires CSI. The scheme has a straightforward encoding
design and low decoding complexity, which can provide full spatial
diversity and requires no CSI \cite{differential_Tarokh_2000,multiple_diff_STBC_2001}.
In this section we restrict the formulation to 2-port Tag antennas,
for clarity of explanation, so that $M=2$ but the method can be applied
to $M=2,4,8$ similarly to coherent OSTBC. 

\subsection{Encoding Algorithm}

In differential OSTBC \cite{differential_Tarokh_2000,multiple_diff_STBC_2001},
the Tag transmits the signal as blocks. The two bits $b_{2i-1}$ and
$b_{2i}$ are mapped to two symbols $u_{2i-1},u_{2i}\in\left\{ \pm1\right\} $,
which are then encoded into the $i$th block $\mathbf{X}_{i}$ as
shown in \eqref{eq:encoding_OSTBC}.

The Tag begins the transmission by sending the first block $\mathbf{X}_{1}$
where $u_{1}$ and $u_{2}$ are arbitrary symbols. The symbols $u_{1}$
and $u_{2}$ are unknown to the Reader and do not convey any information.
The Tag subsequently transmits the signal block in an inductive manner.
Supposing $\mathbf{X}_{i}$ has been sent already, to transmit the
next block $\mathbf{X}_{i+1}$, we need to map $b_{2i+1}$ and $b_{2i+2}$
to symbols $u_{2i+1}$ and $u_{2i+2}$. This mapping depends on $\mathbf{X}_{i}$,
which is different from coherent OSTBC where $\mathbf{X}_{i}$ and
$\mathbf{X}_{i+1}$ are independent of each other. Specifically, we
first use a mapping $\mathcal{M}$ to map $b_{2i+1}$ and $b_{2i+2}$
to a vector $\left[B_{2i+1},B_{2i+2}\right]$, i.e. $\mathcal{M}\left(b_{2i+1},b_{2i+2}\right)=\left[B_{2i+1},B_{2i+2}\right]$
where the mapping is given by \cite{differential_Tarokh_2000}, 
\begin{equation}
\begin{array}{cc}
\mathcal{M}\left(0,0\right)=\left[1,0\right], & \mathcal{M}\left(0,1\right)=\left[0,-1\right]\\
\mathcal{M}\left(1,0\right)=\left[0,1\right], & \mathcal{M}\left(1,1\right)=\left[-1,0\right]
\end{array}.
\end{equation}
The two symbols $u_{2i+1}$ and $u_{2i+2}$ are then computed by 
\begin{equation}
\left[\begin{array}{c}
u_{2i+1}\\
u_{2i+2}
\end{array}\right]=B_{2i+1}\left[\begin{array}{c}
u_{2i-1}\\
u_{2i}
\end{array}\right]+B_{2i+2}\left[\begin{array}{c}
-u_{2i}\\
u_{2i-1}
\end{array}\right].
\end{equation}
Finally, the two symbols $u_{2i+1}$ and $u_{2i+2}$ are encoded into
the block $\mathbf{X}_{i+1}$ as shown in \eqref{eq:encoding_OSTBC}.
This process is inductively repeated until the end of the transmission.
\begin{center}
\begin{figure}[t]
\centering{}\includegraphics[scale=0.36]{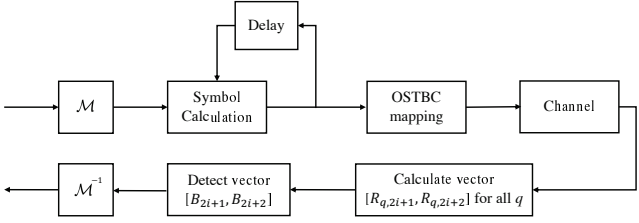}\caption{\label{fig:diff encoder}Block diagram for our proposed non-coherent
OSTBC systems.}
\end{figure}
\par\end{center}

\subsection{Non-Coherent Detection}

For the non-coherent detector, it is not possible to use the accurate
AmBC channel model \eqref{eq:channel model} because of 1) the channel
is nonlinear, and 2) the noise is dependent on the signals and channels.
Therefore, we again leverage the linearized and normalized MIMO channel
model \eqref{eq:Full MIMO} and apply the differential OSTBC detector
\cite{differential_Tarokh_2000}.

Using the differential OSTBC detector, we first detect the vector
$\left[B_{2i+1},B_{2i+2}\right]$ 
\begin{align}
\left[\tilde{B}_{2i+1},\tilde{B}_{2i+2}\right] & =\underset{\mathit{\left[B_{\mathrm{2\mathit{i}+1}},B_{\mathrm{2\mathit{i}+2}}\right]\mathrm{\in\mathcal{V}}}}{\textrm{argmin}}\stackrel[\mathit{q}=1]{\mathit{Q}}{\sum}\left\Vert \left[B_{2i+1},B_{2i+2}\right]\right.\nonumber \\
 & \qquad\qquad\qquad\left.-\left[\mathcal{R}_{q,2i+1},\mathcal{R}_{q,2i+2}\right]\right\Vert ^{2},\label{diff_detect}
\end{align}
where $\mathcal{V}$ denotes the set consisting all the vectors $\left[B_{2i+1},B_{2i+2}\right]$
and $\mathcal{R}_{q,2i+1}$ and $\mathcal{R}_{q,2i+2}$ are 
\begin{equation}
\mathcal{R}_{q,2i+1}=y_{q,2i+1}y_{q,2i-1}+y_{q,2i+2}y_{q,2i},
\end{equation}
\begin{equation}
\mathcal{R}_{q,2i+2}=y_{q,2i+1}y_{q,2i}-y_{q,2i+2}y_{q,2i-1},
\end{equation}
where $y_{q,2i-1}$, $y_{q,2i}$, $y_{q,2i+1}$, and $y_{q,2i+2}$
denote the signals received by the Reader at four consecutive symbol
periods $2i-1$, $2i$, $2i+1$, $2i+2$. We can then detect the two
bits $b_{2i+1}$ and $b_{2i+2}$ from $\left[B_{2i+1},B_{2i+2}\right]$
since they have one-to-one correspondence. The block diagram of the
differential OSTBC system can be found in Fig. \ref{fig:diff encoder}.

\section{Simulation Results}

We provide simulation results for the proposed AmBC system using coherent
and non-coherent OSTBC. A conventional SISO AmBC system ($M=1$, $Q=1$)
using coherent detection \cite{SISO_Lu_2015} and two SISO AmBC systems
using differential detectors \cite{Detection_perfprmance_Wang_2016,Nocoherent_Qian_2017}
are also simulated as benchmarks for comparison. In the simulation,
we assume all the small-scale channel fading $h_{q}^{\mathrm{SR}}$,
$h_{q}^{\mathrm{TR}}$ and $h_{q}^{\mathrm{ST}}$ are distributed
as $\mathcal{CN}\left(0,1\right)$. The hardware implementation loss
by the Tag, $\alpha$, is set as $1.1$ dB \cite{passiveWIFI_Kellogg_2016}.
The Monte Carlo method is used to find the BER.

In the simulation results we focus on the configuration in which the
number of Tag antennas is $M=2.$ This ensures a multiplexing factor
of unity (so as not sacrifice throughput) as well as allowing the
Tag to be compact. Compact 2-port Tag antennas can be easily found
\cite{2port_1,2port_2} but for $M>2$ multi-port antenna designs
with the same density of antennas (in terms of ports per square or
cubic wavelength) are not as straightforward to produce. We also provide
limited results for $M>2$ to demonstrate the versatility of the proposed
method as well as the limitations of the proposed linear AmBC MIMO
channel model.

\subsection{Coherent OSTBC}

\begin{figure}[t]
\begin{centering}
\includegraphics[scale=0.429]{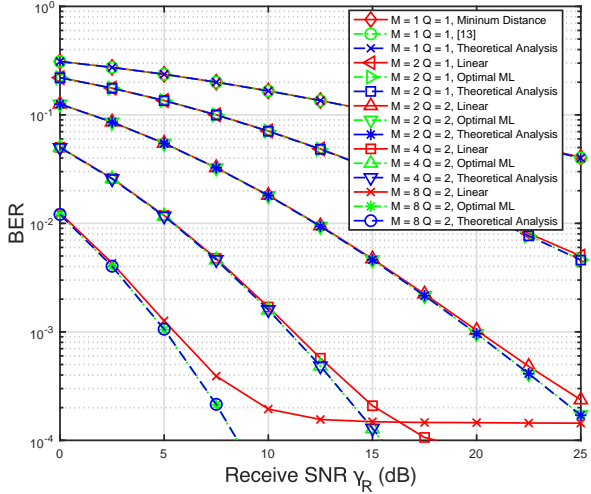}
\par\end{centering}
\begin{centering}
{\footnotesize{}(a) $\Delta\gamma=40$ dB}{\footnotesize\par}
\par\end{centering}
\begin{centering}
\includegraphics[scale=0.429]{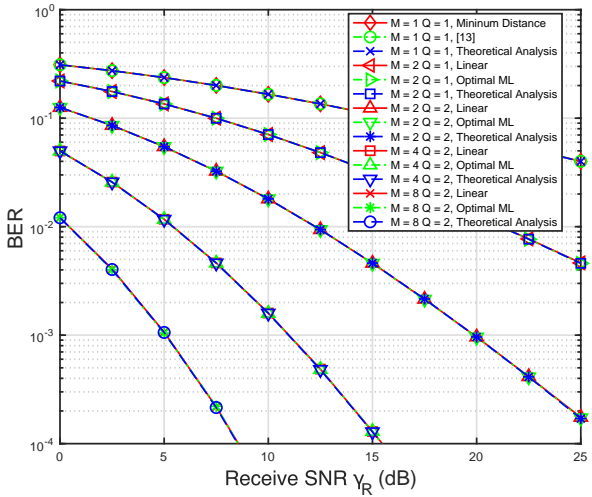}
\par\end{centering}
\begin{centering}
{\footnotesize{}(b) $\Delta\gamma=50$ dB}{\footnotesize\par}
\par\end{centering}
\caption{\label{coherent_detectors}Simulated BERs versus receive SNR $\gamma_{R}$
of the proposed AmBC system using the OSTBC and the SISO AmBC system
using BPSK for $\gamma_{d}=15$ dB at (a) $\Delta\gamma=40$ dB and
(b) $\Delta\gamma=50$ dB.}
\end{figure}

For the proposed AmBC system, we simulate the BER versus receive SNR
$\gamma_{R}$ by using the optimal ML detector \eqref{eq:ML detector}
and the linear detector \eqref{eq:linear detector}. For the SISO
AmBC system, we simulate the BER by using the optimal ML detector
previously proposed \cite{SISO_Lu_2015} and the minimum distance
detector \cite{goldsmith_2005} based on the linearized and normalized
channel. The simulated BERs of the MIMO AmBC system using OSTBC and
the SISO AmBC system using BPSK for the relative SNR $\Delta\gamma=40$
dB and the direct link SNR $\gamma_{d}=15$ dB, which are practical
values, are shown in Fig. \ref{coherent_detectors}. Recall that the
relative SNR $\Delta\gamma$ depends on the hardware implementation
loss $\alpha$ and path loss $A_{\mathrm{TR}}$ while the direct link
SNR $\gamma_{d}$ is $P_{s}/\sigma^{2}$. Besides, $\gamma_{R}$ depends
on $\Delta\gamma$, $\gamma_{d}$, $N$, and the channels as shown
in \eqref{eq:yr-siso}. We use \eqref{eq:PSISO} and \eqref{eq:PAlamouti}
to calculate the theoretical BER for a given channel realization.
Using Monte Carlo method, we generate multiple independent channel
realizations and average the obtained theoretical BERs. The averaged
BER results are denoted as theoretical analysis and included in Fig.
\ref{coherent_detectors} as a reference.

In Fig. \ref{coherent_detectors}, five configurations are shown:
$1\times1$, $2\times1$, $2\times2$, $4\times2$ and $8\times2$.
In terms of approximation accuracy, comparing the five groups of curves,
three key observations can be highlighted.

\textit{First}, from Fig. \ref{coherent_detectors}(a) and (b), when
$M=1,2$, the optimal ML detector has almost the same BER performance
as the linear detector or the minimum distance detector, which shows
that our proposed linearized and normalized MIMO channel \eqref{eq:Full MIMO}
well approximates the accurate AmBC channel \eqref{eq:channel model}.
The linear detector based on the proposed channel model however has
a much lower computational complexity than the optimal ML detector.
Therefore, it is beneficial to use our proposed linearized and normalized
MIMO channel. It can be seen that the BER performance of the linear
or minimum distance detector matches well with the theoretical analysis
results \eqref{eq:PSISO} and \eqref{eq:PAlamouti}, which validates
the correctness of the analysis.

\textit{Second}, from Fig. \ref{coherent_detectors}(a), when $M=4$,
the performance gap becomes a bit larger and when $M=8$, the performance
gap is obvious and there is an error floor, which indicates that the
BER cannot decrease with the SNR without a limitation. Our approximation
does not work well in these cases. The error floor is due to the approximation
of the signal to a linear function by omitting the quadratic term
$P_{s}\left|\mathbf{h}_{q}^{\mathrm{TR}}\mathbf{G}\mathbf{x}\right|^{2}$.
When increasing the antenna number, the omitted term $P_{s}\left|\mathbf{h}_{q}^{\mathrm{TR}}\mathbf{G}\mathbf{x}\right|^{2}$
causes higher error and affects the accuracy in the minimum distance
detector so that it causes an error floor.

\textit{Third}, comparing Fig. \ref{coherent_detectors}(a) with (b),
we can find that the performance gap between the optimal ML detector
and the linear detector is eliminated for $M=1,2,4,8$ when $\Delta\gamma$
increases from 40 to 50 dB, which shows that higher $\Delta\gamma$
makes the approximations more accurate even for larger Tag antenna
numbers. In addition, $\Delta\gamma$ increases with the distance
between Tag and Reader, as will be shown in the following. In other
words, as we increase the distance between Tag and Reader, our proposed
approximations become more accurate even for more Tag antennas.

In summary we can conclude, that our proposed channel model approximations
becomes less accurate when we increase the number of antennas and
on the contrary become more accurate when we increase $\Delta\gamma$
or equivalently the distance between Tag and Reader. 

\begin{figure}[t]
\begin{centering}
\includegraphics[scale=0.43]{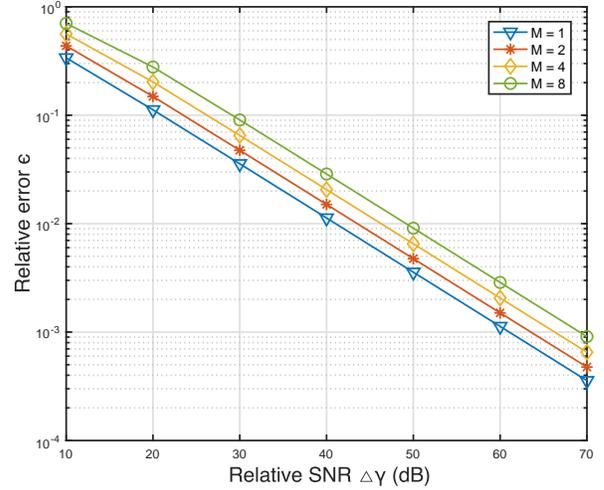}
\par\end{centering}
\caption{\label{approximation error}The relative error for omitting $P_{s}\left|\mathbf{h}_{q}^{\mathrm{TR}}\mathbf{G}\mathbf{x}\right|^{2}$
in the information signal versus $\Delta\gamma$ with different number
of antennas}
\end{figure}

Fig. \ref{approximation error} shows the relative error ratio $\epsilon$
\eqref{eq: approximation error} for omitting $P_{s}\left|\mathbf{h}_{q}^{\mathrm{TR}}\mathbf{G}\mathbf{x}\right|^{2}$
in the information signal versus $\Delta\gamma$ with different number
of antennas. We can find that the relative error increases with the
Tag antenna number and decreases with $\Delta\gamma$, which agrees
well with the above results.

To conclude, at $\Delta\gamma=40$ dB, our channel model approximation
is accurate for small numbers of Tag antennas ($M\leq2$) which satisfies
the size constraint for practical Tag configurations. If we extend
the communication distance between the Tag and Reader to increase
the relative SNR e.g., $\Delta\gamma=50$ dB, the approximation can
be more accurate even for more Tag antennas such as $M=4,8$. Furthermore,
to emphasize the benefits of implementing multiple antennas at the
Tag as well as Reader two advantages need to be highlighted:

\textit{First}, it is important to note that the total transmit power
in the MIMO AmBC system linearly increases with $M$, which is different
from conventional MIMO communication where the total transmit power
is fixed. If we set the transmit symbols as $u_{Mi-M+1},...,u_{Mi}\in\left\{ \pm1/\sqrt{M}\right\} $
as adopted in the conventional MIMO system, the BER curves for $M=2,4,8$
in Fig. \ref{coherent_detectors} will right shift by $10\mathrm{lg}M$
dB. Therefore, using more Tag antennas has the benefit of power gain,
which effectively increases the power of the backscattered signal
so as to decrease the BER, and this is unique in the AmBC setting.

\textit{Second}, comparing the BER curves of $M=1$, $Q=1$ and $M=2$,
$Q=1$, we can find that the slope of the BER curve is doubled by
using two Tag antennas. The enhanced signal detection performance
arises from leveraging the Tag diversity gain. Furthermore, comparing
the BER curves of $M=2$, $Q=1$ and $M=2$, $Q=2$, we can find that
the slope of the BER curve is doubled by using two Reader antennas.
The enhanced signal detection performance arises from leveraging the
Reader diversity gain. Overall, compared with our benchmark \cite{SISO_Lu_2015},
our proposed approach shows the benefit of using OSTBC in the multiple-antenna
Tag multiple-antenna Reader AmBC system to enhance the signal detection
performance. More generally, a diversity gain of $MQ$ can be achieved
by using more Tag/Reader antennas in AmBC.

\begin{figure}[t]
\centering{}\includegraphics[scale=0.43]{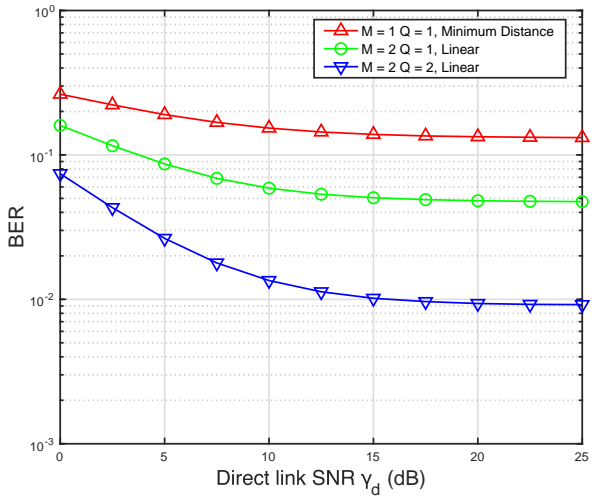}\caption{\label{direct_link} Simulated BERs versus direct link SNR $\gamma_{d}$
of the proposed AmBC system using coherent OSTBC and the SISO AmBC
system using BPSK for $\Delta\gamma=40$ dB and $N=40000$.}
\end{figure}

We also study the effect of direct link SNR $\gamma_{d}$ on the BER
performance of the MIMO AmBC system and the SISO AmBC system. The
simulated BERs versus $\gamma_{d}$ are shown in Fig. \ref{direct_link}.
We only consider using the linear detector and the minimum distance
detector since they have almost the same BERs as the optimal ML detector.
From Fig. \ref{direct_link}, we can observe that the BERs decrease
with $\gamma_{d}$ at fixed $N=40000$ for the three configurations.
This is because the receive SNR $\gamma_{R}$ increases with $\gamma_{d}$.
Therefore, high $\gamma_{d}$ is beneficial for AmBC to decrease the
BER. However, BER cannot be decreased without limit by simply increasing
$\gamma_{d}$. As shown in Fig. \ref{direct_link}, the BERs saturate
when $\gamma_{d}$ is large for the three configurations. This is
because when $\gamma_{d}\rightarrow\infty$, $\gamma_{R}$ saturates
to $\frac{4N}{\Delta\gamma}\mathbb{E}\left[\mathfrak{R}^{2}\left\{ h_{1,1}^{\mathrm{TR}}h_{1}^{\mathrm{ST}}/h_{1}^{\mathrm{SR}}\right\} \right]$
as shown in Section III.C. In addition, we can observe that using
more Tag/Reader antennas can decrease the BER given the same $\gamma_{d}$,
which again shows the benefit of multiple Tag/Reader antennas.

\begin{figure}[t]
\begin{centering}
\includegraphics[scale=0.43]{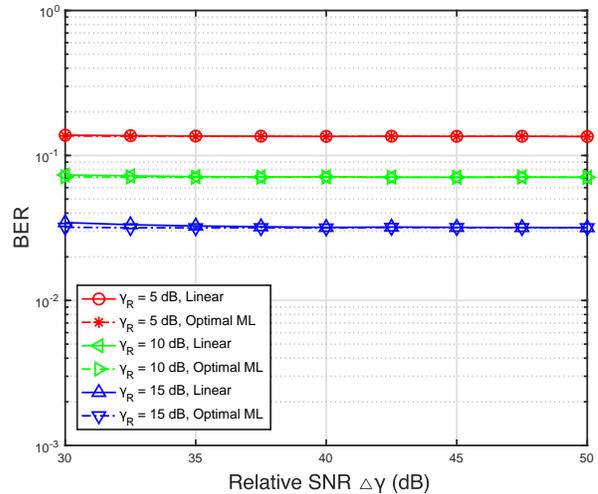}
\par\end{centering}
\centering{}\caption{\label{relative_SNR}Simulated BERs versus relative SNR $\Delta\gamma$
of the proposed AmBC system using coherent OSTBC for $M=2$, $Q=1$
at the receive SNRs $\gamma_{R}=$ 5, 10, 15 dB.}
\end{figure}

Because our proposed linearized and normalized MIMO channel \eqref{eq:Full MIMO}
is based on the relative SNR $\Delta\gamma$ being high in AmBC, we
study the effect of $\Delta\gamma$ on the BER performance. The simulated
BERs versus $\Delta\gamma$ of the proposed AmBC ($M=2$, $Q=1$)
at the receive SNRs $\gamma_{R}=$ 5, 10, 15 dB are shown in Fig.
\ref{relative_SNR}. From Fig. \ref{relative_SNR}, we can observe
that given a fixed $\gamma_{R}$, the optimal ML detector and the
linear detector have almost the same BERs which do not change with
$\Delta\gamma$ when $\Delta\gamma$ is larger than 35 dB. However,
when $\Delta\gamma$ is small, e.g. at 30 dB, there is a small BER
performance gap between the optimal ML detector and the linear detector,
which implies that the linearized and normalized MIMO channel model
has some approximation error. Fortunately, $\Delta\gamma$ in AmBC
is usually higher than 30 dB.

\begin{figure}[t]
\centering{}\includegraphics[scale=0.43]{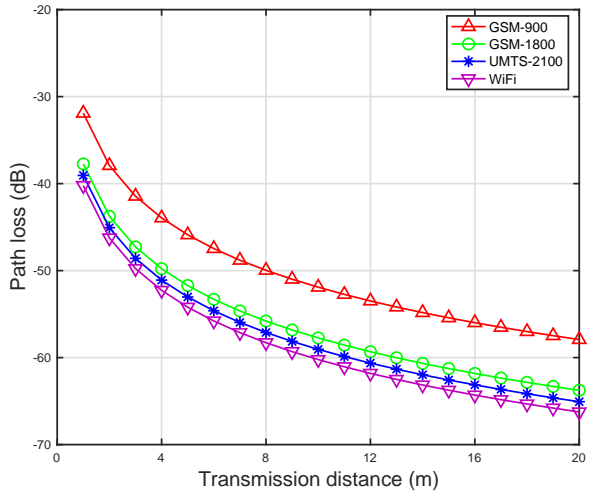}\caption{\label{path loss}Path loss versus transmission distance between the
Tag and Reader for ambient RF signals from different wireless systems.}
\end{figure}

The relative SNR $\Delta\gamma$ has a one-to-one correspondence with
the path loss since $\Delta\gamma=1/\alpha^{2}A_{\mathrm{TR}}^{2}$.
Therefore to show the range of $\Delta\gamma$, we use the Friis equation
\cite{goldsmith_2005} to estimate the path loss for the ambient RF
signals from different wireless systems. The path loss versus the
transmission distance between the Tag and Reader for the ambient RF
signals of different wireless systems is shown in Fig. \ref{path loss}
when we use a path-loss exponent of two. We consider common ambient
RF signals from wireless systems including GSM-900 (925-960 MHz),
GSM-1800 (1.805-1.88 GHz), UMTS-2100 (2.11-2.17 GHz) and WiFi (2.4-2.48
GHz), which have been well exploited for ambient RF energy harvesting
\cite{ShanpuShen2019_TMTT_Freqdepend,ShanpuShen2019_TIE_HybridCombining}.
It should be noted that in Fig. \ref{path loss}, the path loss decreases
with the transmission distance and it is below $-30$ dB when the
transmission distance is larger than 1 meter for all considered ambient
RF signals. In other words, $\Delta\gamma$ is at least greater than
30 dB when the Tag and Reader are separated by 1 meter or more. In
previous simulations, we set $\Delta\gamma=40$ dB so the communication
distance between Tag and Reader is around 1-2 meter, which is a practical
set up. Since Fig. \ref{relative_SNR} shows that our proposed linearized
and normalized MIMO channel model is accurate when the relative SNR
$\Delta\gamma$ is larger than 30 dB, or equivalently the distance
between the Tag and Reader is longer than 1 meter, the proposed channel
model is shown valid for practical AmBC system configurations and
based on that the linear detector provides the optimal BER performance
($M\leq2$).

\subsection{Non-Coherent OSTBC}

For the MIMO AmBC system, we simulate the BER by using the differential
OSTBC detector \eqref{diff_detect}. For the SISO AmBC system, we
reproduce the methods previously proposed \cite{Detection_perfprmance_Wang_2016,Nocoherent_Qian_2017}
which also implement differential codes in an AmBC system.

\begin{figure}[t]
\begin{centering}
\includegraphics[scale=0.43]{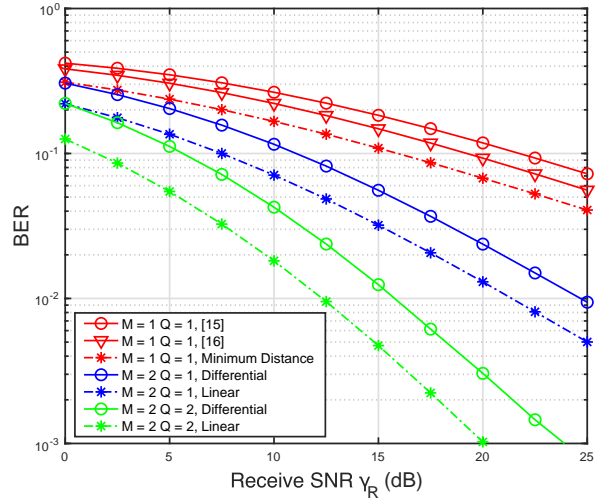}
\par\end{centering}
\centering{}\caption{\label{diff}Simulated BERs versus receive SNR $\gamma_{R}$ of the
proposed AmBC system using the differential OSTBC and the SISO AmBC
system using 2DPSK for $\Delta\gamma=40$ dB and $\gamma_{d}=15$
dB.}
\end{figure}

The simulated BERs of the proposed AmBC system using our differential
OSTBC and the SISO AmBC system benchmark with relative SNR $\Delta\gamma$
= $40$ dB and direct link SNR $\gamma_{d}=15$ dB are shown in Fig.
\ref{diff}. Comparing Fig. \ref{coherent_detectors} with Fig. \ref{diff},
we can make three key observations.

\textit{First}, compared with coherent OSTBC, using differential OSTBC
can achieve the same slope in BER, but there is a constant BER performance
loss, which results from not requiring any CSI. For SISO AmBC, there
is also a constant BER loss between the minimum distance detector
in BPSK and differential detector previously proposed \cite{Detection_perfprmance_Wang_2016,Nocoherent_Qian_2017}
for the same reason. It can also be observed that the performance
loss between minimum distance detector in BPSK and \cite{Nocoherent_Qian_2017}
is the same as $M=2$, $Q=1$, and $M=2$, $Q=2$ cases while the
gap between minimum distance detector in BPSK and \cite{Detection_perfprmance_Wang_2016}
is larger than other cases. The reason is that some information is
lost during energy subtraction \cite{Detection_perfprmance_Wang_2016}.

\textit{Second}, similar to the coherent case, using more Tag antennas
has the benefit of power gain, which can effectively increase the
power of the backscattered signal so as to decrease the BER.

\textit{Third}, comparing the BER curves of $M=1$, $Q=1$, $M=2$,
$Q=1$, and $M=2$, $Q=2$, we find that the slope of the BER curve
can be doubled by using two Tag antennas or Reader antennas. Therefore,
using differential OSTBC can leverage the Tag/Reader diversity gain
and enhance the signal detection performance without the requirement
of any CSI.

\begin{figure}[t]
\begin{centering}
\includegraphics[scale=0.43]{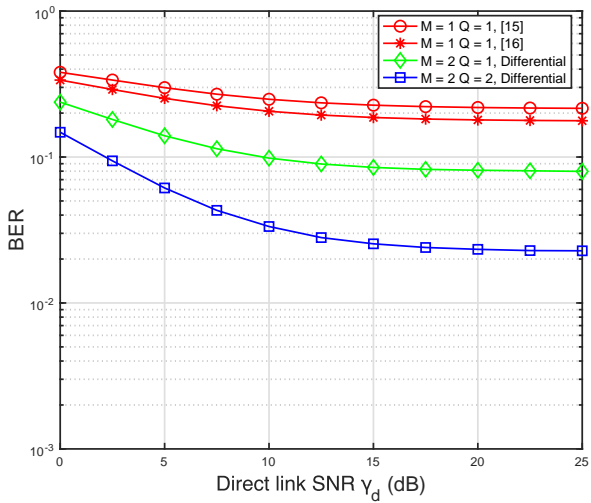}
\par\end{centering}
\begin{centering}
{\footnotesize{}(a)}{\footnotesize\par}
\par\end{centering}
\begin{centering}
\includegraphics[scale=0.43]{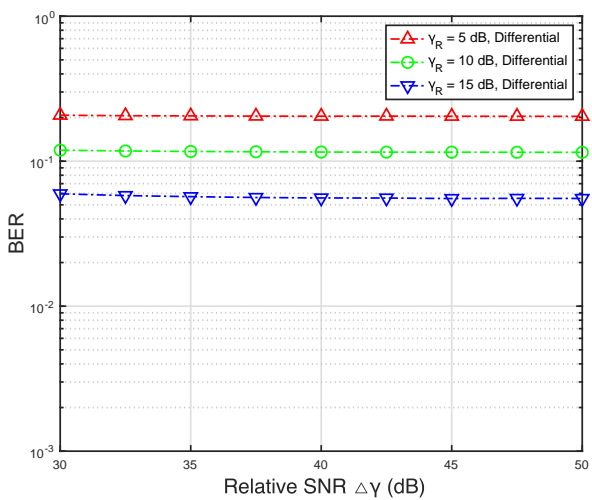}
\par\end{centering}
\begin{centering}
{\footnotesize{}(b)}{\footnotesize\par}
\par\end{centering}
\centering{}\caption{\label{diff_2}(a) Simulated BERs versus direct link SNR $\gamma_{d}$
of the proposed AmBC system using differential OSTBC and the SISO
AmBC system \cite{Detection_perfprmance_Wang_2016}, \cite{Nocoherent_Qian_2017}
for $\Delta\gamma=40$ dB and $N=40000$. (b) Simulated BERs versus
relative SNR $\Delta\gamma$ of the proposed AmBC system using differential
OSTBC for $M=2$, $Q=1$ at the receive SNRs $\gamma_{R}=$ 5, 10,
15 dB.}
\end{figure}

Similar to the coherent case, we also show how the direct link SNR
$\gamma_{d}$ and the relative SNR $\Delta\gamma$ affect the BER
performance of the proposed AmBC system using differential OSTBC as
shown in Fig. \ref{diff_2}(a) and Fig. \ref{diff_2}(b), respectively.
We can also make observations that are similar to the case of using
coherent OSTBC, and these are summarized as 1) BERs decrease with
$\gamma_{d}$ and saturate to a constant when $\gamma_{d}\rightarrow\infty$;
2) BERs do not change with $\Delta\gamma$ given a fixed $\gamma_{R}$
when $\Delta\gamma$ is larger than 35 dB, and the BERs have little
increase when $\Delta\gamma$ is at around 30 dB. The only difference
between the coherent and non-coherent cases is that there is a constant
BER performance loss by using differential OSTBC. However it comes
at the advantage of not requiring any CSI.

\section{Conclusion}

We have proposed the use of multiple-antenna Tag and multiple-antenna
Reader in AmBC systems by applying coherent and non-coherent OSTBC
to leverage both diversity and power gain. Both codes do not require
CSIT and the non-coherent OSTBC do not require CSIR either. The multiple-antenna
Tag linearly increases (linear with the number of Tag antennas) the
total backscattered signal power and this is different from conventional
MIMO where transmit power is fixed. Furthermore, the proposed approach
is compatible with using multiple-antenna Reader to further improve
diversity gain.

We have also derived an accurate channel model for the AmBC system
with a multiple-antenna Tag and a multiple-antenna Reader by using
the averaging mechanism. The accurate AmBC channel model is challenging
to use since it is nonlinear with noise that is dependent on the signals
and the channels and cannot be used with OSTBC. To overcome the challenges
of using the accurate AmBC channel model, we have proposed a linearized
and normalized MIMO channel model. The channel model is accurate when
there are two antennas at the Tag but becomes less accurate when the
number of antennas is increased. On the contrary the model becomes
more accurate when $\Delta\gamma$ is increased or equivalently the
distance between Tag and Reader is increased.

Two coherent detectors, the optimal ML detector and the linear detector
have been investigated and a closed-form solution for the BER was
also provided. In addition a non-coherent detector based on differential
OSTBC is also investigated. Using this there is no need for CSI and
the AmBC system design complexity and power consumption can be reduced.
The corresponding differential OSTBC detector based on the linearized
and normalized MIMO channel was also provided.

Simulation results demonstrate that the proposed MIMO AmBC system
using coherent and non-coherent OSTBC has performance better than
the conventional SISO AmBC system. The simulation results show that
using multiple Tag antennas can increase the total power of the backscattered
signal. It is also shown that using OSTBC and differential OSTBC in
the proposed AmBC system can effectively enhance BER performance compared
compared with SISO AmBC systems.


\begin{thebibliography}{10}
	\providecommand{\url}[1]{#1}
	\csname url@samestyle\endcsname
	\providecommand{\newblock}{\relax}
	\providecommand{\bibinfo}[2]{#2}
	\providecommand{\BIBentrySTDinterwordspacing}{\spaceskip=0pt\relax}
	\providecommand{\BIBentryALTinterwordstretchfactor}{4}
	\providecommand{\BIBentryALTinterwordspacing}{\spaceskip=\fontdimen2\font plus
		\BIBentryALTinterwordstretchfactor\fontdimen3\font minus
		\fontdimen4\font\relax}
	\providecommand{\BIBforeignlanguage}[2]{{%
			\expandafter\ifx\csname l@#1\endcsname\relax
			\typeout{** WARNING: IEEEtran.bst: No hyphenation pattern has been}%
			\typeout{** loaded for the language `#1'. Using the pattern for}%
			\typeout{** the default language instead.}%
			\else
			\language=\csname l@#1\endcsname
			\fi
			#2}}
	\providecommand{\BIBdecl}{\relax}
	\BIBdecl
	
	\bibitem{reflected_1948_stockman}
	H.~{Stockman}, ``Communication by means of reflected power,'' \emph{Proc. IRE},
	vol.~36, no.~10, pp. 1196--1204, 1948.
	
	\bibitem{history_rfid_Landt_2005}
	J.~{Landt}, ``The history of {RFID},'' \emph{IEEE Potentials}, vol.~24, no.~4,
	pp. 8--11, 2005.
	
	\bibitem{bistatic_Kimionis_2013}
	J.~{Kimionis}, A.~{Bletsas}, and J.~N. {Sahalos}, ``Bistatic backscatter radio
	for power-limited sensor networks,'' in \emph{2013 IEEE Global Communications
		Conference (GLOBECOM)}, 2013, pp. 353--358.
	
	\bibitem{MIMO_RFID_Boyer_2014}
	C.~{Boyer} and S.~{Roy}, ``Backscatter communication and {RFID}: Coding,
	energy, and {MIMO} analysis,'' \emph{IEEE Trans. Commun.}, vol.~62, no.~3,
	pp. 770--785, 2014.
	
	\bibitem{survery_VanHuynh_2018}
	N.~{Van Huynh}, D.~T. {Hoang}, X.~{Lu}, D.~{Niyato}, P.~{Wang}, and D.~I.
	{Kim}, ``Ambient backscatter communications: A contemporary survey,''
	\emph{IEEE Communications Surveys Tutorials}, vol.~20, no.~4, pp. 2889--2922,
	2018.
	
	\bibitem{thin_Liu_2013}
	V.~Liu, A.~Parks, V.~Talla, S.~Gollakota, D.~Wetherall, and J.~R. Smith,
	``Ambient backscatter: Wireless communication out of thin air,'' \emph{ACM
		SIGCOMM Computer Communication Review}, vol.~43, no.~4, pp. 39--50, 2013.
	
	\bibitem{FMbackscatter_Wang_2017}
	\BIBentryALTinterwordspacing
	A.~Wang, V.~Iyer, V.~Talla, J.~R. Smith, and S.~Gollakota, ``{FM} backscatter:
	Enabling connected cities and smart fabrics,'' in \emph{14th {USENIX}
		Symposium on Networked Systems Design and Implementation ({NSDI} 17)}.\hskip
	1em plus 0.5em minus 0.4em\relax Boston, MA: {USENIX} Association, Mar. 2017,
	pp. 243--258. [Online]. Available:
	\url{https://www.usenix.org/conference/nsdi17/technical-sessions/presentation/wang-anran}
	\BIBentrySTDinterwordspacing
	
	\bibitem{AmBC_FM_Daskalakis_2017}
	S.~N. {Daskalakis}, J.~{Kimionis}, A.~{Collado}, G.~{Goussetis}, M.~M.
	{Tentzeris}, and A.~{Georgiadis}, ``Ambient backscatterers using {FM}
	broadcasting for low cost and low power wireless applications,'' \emph{IEEE
		Trans. Microw. Theory Techn.}, vol.~65, no.~12, pp. 5251--5262, 2017.
	
	\bibitem{WiFiback_Kwllogg_2014}
	B.~Kellogg, A.~Parks, S.~Gollakota, J.~R. Smith, and D.~Wetherall, ``{W}i-{F}i
	backscatter: Internet connectivity for rf-powered devices,'' in
	\emph{Proceedings of the 2014 ACM Conference on SIGCOMM}, 2014, pp. 607--618.
	
	\bibitem{Backfi_Bharadia_2015}
	D.~Bharadia, K.~R. Joshi, M.~Kotaru, and S.~Katti, ``Backfi: High throughput
	{W}i{F}i backscatter,'' \emph{ACM SIGCOMM Computer Communication Review},
	vol.~45, no.~4, pp. 283--296, 2015.
	
	\bibitem{passiveWIFI_Kellogg_2016}
	\BIBentryALTinterwordspacing
	B.~Kellogg, V.~Talla, S.~Gollakota, and J.~R. Smith, ``Passive {W}i-{F}i:
	Bringing low power to {W}i-{F}i transmissions,'' in \emph{13th {USENIX}
		Symposium on Networked Systems Design and Implementation ({NSDI} 16)}.\hskip
	1em plus 0.5em minus 0.4em\relax Santa Clara, CA: {USENIX} Association, Mar.
	2016, pp. 151--164. [Online]. Available:
	\url{https://www.usenix.org/conference/nsdi16/technical-sessions/presentation/kellogg}
	\BIBentrySTDinterwordspacing
	
	\bibitem{Efficient_BackFi_Ji_2019}
	B.~{Ji}, B.~{Xing}, K.~{Song}, C.~{Li}, H.~{Wen}, and L.~{Yang}, ``The
	efficient {B}ack{F}i transmission design in ambient backscatter communication
	systems for {I}o{T},'' \emph{IEEE Access}, vol.~7, pp. 31\,397--31\,408,
	2019.
	
	\bibitem{SISO_Lu_2015}
	K.~{Lu}, G.~{Wang}, F.~{Qu}, and Z.~{Zhong}, ``Signal detection and ber
	analysis for {RF}-powered devices utilizing ambient backscatter,'' in
	\emph{2015 International Conference on Wireless Communications Signal
		Processing (WCSP)}, 2015, pp. 1--5.
	
	\bibitem{Exact_BER_Devineni_2019}
	J.~K. {Devineni} and H.~S. {Dhillon}, ``Ambient backscatter systems: Exact
	average bit error rate under fading channels,'' \emph{IEEE Transactions on
		Green Communications and Networking}, vol.~3, no.~1, pp. 11--25, 2019.
	
	\bibitem{Detection_perfprmance_Wang_2016}
	G.~{Wang}, F.~{Gao}, R.~{Fan}, and C.~{Tellambura}, ``Ambient backscatter
	communication systems: Detection and performance analysis,'' \emph{IEEE
		Trans. Commun.}, vol.~64, no.~11, pp. 4836--4846, 2016.
	
	\bibitem{Nocoherent_Qian_2017}
	J.~{Qian}, F.~{Gao}, G.~{Wang}, S.~{Jin}, and H.~{Zhu}, ``Noncoherent
	detections for ambient backscatter system,'' \emph{IEEE Trans. Wireless
		Commun.}, vol.~16, no.~3, pp. 1412--1422, 2017.
	
	\bibitem{MPSK_Qian_2019}
	J.~{Qian}, A.~N. {Parks}, J.~R. {Smith}, F.~{Gao}, and S.~{Jin}, ``{I}o{T}
	communications with ${M}$-{PSK} modulated ambient backscatter: Algorithm,
	analysis, and implementation,'' \emph{IEEE Internet Things J.}, vol.~6,
	no.~1, pp. 844--855, 2019.
	
	\bibitem{MFSK_Tao_2019}
	Q.~{Tao}, C.~{Zhong}, K.~{Huang}, X.~{Chen}, and Z.~{Zhang}, ``Ambient
	backscatter communication systems with {MFSK} modulation,'' \emph{IEEE Trans.
		Wireless Commun.}, vol.~18, no.~5, pp. 2553--2564, 2019.
	
	\bibitem{Swiching_frequency_Vougioukas_2019}
	G.~{Vougioukas} and A.~{Bletsas}, ``Switching frequency techniques for
	universal ambient backscatter networking,'' \emph{IEEE J. Sel. Areas
		Commun.}, vol.~37, no.~2, pp. 464--477, 2019.
	
	\bibitem{matched_filter_Choi_2019}
	J.~{Choi}, ``Matched-filter-based backscatter communication for {I}o{T} devices
	over ambient {OFDM} carrier,'' \emph{IEEE Internet Things J.}, vol.~6, no.~6,
	pp. 10\,229--10\,239, 2019.
	
	\bibitem{Coding_Detection_Liu_2017}
	Y.~{Liu}, G.~{Wang}, Z.~{Dou}, and Z.~{Zhong}, ``Coding and detection schemes
	for ambient backscatter communication systems,'' \emph{IEEE Access}, vol.~5,
	pp. 4947--4953, 2017.
	
	\bibitem{Manchester_coding_Tao_2018}
	Q.~{Tao}, C.~{Zhong}, H.~{Lin}, and Z.~{Zhang}, ``Symbol detection of ambient
	backscatter systems with {M}anchester coding,'' \emph{IEEE Trans. Wireless
		Commun.}, vol.~17, no.~6, pp. 4028--4038, 2018.
	
	\bibitem{Optimal_Resource_Allocation_Full_Duplex_Yang_2019}
	G.~{Yang}, D.~{Yuan}, Y.~{Liang}, R.~{Zhang}, and V.~C.~M. {Leung}, ``Optimal
	resource allocation in full-duplex ambient backscatter communication networks
	for wireless-powered {I}o{T},'' \emph{IEEE Internet Things J.}, vol.~6,
	no.~2, pp. 2612--2625, 2019.
	
	\bibitem{Time_scheduling_Liu_2019}
	X.~{Liu}, Y.~{Gao}, and F.~{Hu}, ``Optimal time scheduling scheme for wireless
	powered ambient backscatter communications in {I}o{T} networks,'' \emph{IEEE
		Internet Things J.}, vol.~6, no.~2, pp. 2264--2272, 2019.
	
	\bibitem{TP_Splitting_Ma_2019}
	Z.~{Ma}, C.~{He}, Y.~{Rao}, J.~{Jiang}, S.~{Ma}, F.~{Gao}, and L.~{Xing},
	``Time- and power-splitting strategies for ambient backscatter system,''
	\emph{IEEE Access}, vol.~7, pp. 40\,068--40\,077, 2019.
	
	\bibitem{Modeling_performance_Cellular_Shi_2020}
	L.~{Shi}, R.~Q. {Hu}, Y.~{Ye}, and H.~{Zhang}, ``Modeling and performance
	analysis for ambient backscattering underlaying cellular networks,''
	\emph{IEEE Trans. Veh. Technol.}, vol.~69, no.~6, pp. 6563--6577, 2020.
	
	\bibitem{Modeling_performance_Darsena_2017}
	D.~{Darsena}, G.~{Gelli}, and F.~{Verde}, ``Modeling and performance analysis
	of wireless networks with ambient backscatter devices,'' \emph{IEEE Trans.
		Commun.}, vol.~65, no.~4, pp. 1797--1814, 2017.
	
	\bibitem{Ergodic_rat_Zhou_2019}
	S.~{Zhou}, W.~{Xu}, K.~{Wang}, C.~{Pan}, M.~{Alouini}, and A.~{Nallanathan},
	``Ergodic rate analysis of cooperative ambient backscatter communication,''
	\emph{IEEE Wireless Commun. Lett.}, vol.~8, no.~6, pp. 1679--1682, 2019.
	
	\bibitem{outage_cooperative_Ding_2020}
	H.~{Ding}, D.~B. {da Costa}, and J.~{Ge}, ``Outage analysis for cooperative
	ambient backscatter systems,'' \emph{IEEE Wireless Commun. Lett.}, vol.~9,
	no.~5, pp. 601--605, 2020.
	
	\bibitem{Outage_performance_Ye_2020}
	Y.~{Ye}, L.~{Shi}, X.~{Chu}, and G.~{Lu}, ``On the outage performance of
	ambient backscatter communications,'' \emph{IEEE Internet Things J.}, vol.~7,
	no.~8, pp. 7265--7278, 2020.
	
	\bibitem{Approximate_BER_Tagselection_Zhou_2017}
	X.~{Zhou}, G.~{Wang}, Y.~{Wang}, and J.~{Cheng}, ``An approximate {BER}
	analysis for ambient backscatter communication systems with tag selection,''
	\emph{IEEE Access}, vol.~5, pp. 22\,552--22\,558, 2017.
	
	\bibitem{Capacity_Tag_delection_Li_2019}
	D.~{Li}, W.~{Peng}, and F.~{Hu}, ``Capacity of backscatter communication
	systems with tag selection,'' \emph{IEEE Trans. Veh. Technol.}, vol.~68,
	no.~10, pp. 10\,311--10\,314, 2019.
	
	\bibitem{Roubust_design_Zhang_2019}
	Y.~{Zhang}, B.~{Li}, F.~{Gao}, and Z.~{Han}, ``A robust design for ultra
	reliable ambient backscatter communication systems,'' \emph{IEEE Internet
		Things J.}, vol.~6, no.~5, pp. 8989--8999, 2019.
	
	\bibitem{New_approach_RF_POWER_Hoang_2017}
	D.~T. {Hoang}, D.~{Niyato}, P.~{Wang}, D.~I. {Kim}, and Z.~{Han}, ``Ambient
	backscatter: A new approach to improve network performance for {RF}-powered
	cognitive radio networks,'' \emph{IEEE Trans. Commun.}, vol.~65, no.~9, pp.
	3659--3674, 2017.
	
	\bibitem{Opportunistic_AmBC_2019}
	R.~{Kishore}, S.~{Gurugopinath}, P.~C. {Sofotasios}, S.~{Muhaidat}, and
	N.~{Al-Dhahir}, ``Opportunistic ambient backscatter communication in
	{RF}-powered cognitive radio networks,'' \emph{IEEE Trans. on Cogn. Commun.
		Netw.}, vol.~5, no.~2, pp. 413--426, 2019.
	
	\bibitem{Hardware_Efficient_detection_Tao_2019}
	Y.~{Tao}, B.~{Li}, C.~{Zhao}, and Y.~{Liang}, ``Hardware-efficient signal
	detection for ambient backscattering communications,'' \emph{IEEE Commun.
		Lett.}, vol.~23, no.~12, pp. 2196--2199, 2019.
	
	\bibitem{umo_Parks_2014}
	A.~N. Parks, A.~Liu, S.~Gollakota, and J.~R. Smith, ``Turbocharging ambient
	backscatter communication,'' \emph{ACM SIGCOMM Computer Communication
		Review}, vol.~44, no.~4, pp. 619--630, 2014.
	
	\bibitem{expand_umo_Ma_2018}
	S.~Ma, G.~Wang, Y.~Wang, and Z.~Zhao, ``Signal ratio detection and approximate
	performance analysis for ambient backscatter communication systems with
	multiple receiving antennas,'' \emph{Mobile Networks and Applications},
	vol.~23, no.~6, pp. 1478--1486, 2018.
	
	\bibitem{eigenvalue_detection_Tao_2019}
	Q.~{Tao}, C.~{Zhong}, X.~{Chen}, H.~{Lin}, and Z.~{Zhang}, ``Maximum-eigenvalue
	detector for multiple antenna ambient backscatter communication systems,''
	\emph{IEEE Trans. Veh. Technol.}, vol.~68, no.~12, pp. 12\,411--12\,415,
	2019.
	
	\bibitem{multiple_antenna_cognitive_Guo_2019}
	H.~{Guo}, Q.~{Zhang}, S.~{Xiao}, and Y.~{Liang}, ``Exploiting multiple antennas
	for cognitive ambient backscatter communication,'' \emph{IEEE Internet Things
		J.}, vol.~6, no.~1, pp. 765--775, 2019.
	
	\bibitem{Constellation_Learning_Based_Zhang_2019}
	Q.~{Zhang}, H.~{Guo}, Y.~{Liang}, and X.~{Yuan}, ``Constellation learning-based
	signal detection for ambient backscatter communication systems,'' \emph{IEEE
		J. Sel. Areas Commun.}, vol.~37, no.~2, pp. 452--463, 2019.
	
	\bibitem{frequency_diverse_array_Hu_2020}
	Y.~Q. {Hu}, H.~{Chen}, S.~L. {Ji}, and W.~Q. {Wang}, ``Ambient backscatter
	communication with frequency diverse array for enhanced channel capacity and
	detection performance,'' \emph{IEEE Sensors J.}, vol.~20, no.~18, pp.
	10\,876--10\,885, 2020.
	
	\bibitem{Cooperative_YCLiang_2018}
	G.~{Yang}, Q.~{Zhang}, and Y.~{Liang}, ``Cooperative ambient backscatter
	communications for green internet-of-things,'' \emph{IEEE Internet Things
		J.}, vol.~5, no.~2, pp. 1116--1130, 2018.
	
	\bibitem{MRC_Adaptive_Li_2018}
	D.~{Li} and Y.~{Liang}, ``Adaptive ambient backscatter communication systems
	with {MRC},'' \emph{IEEE Trans. Veh. Technol.}, vol.~67, no.~12, pp.
	12\,352--12\,357, 2018.
	
	\bibitem{OFDM_multiple_receive_Yang_2018}
	G.~{Yang}, Y.~{Liang}, R.~{Zhang}, and Y.~{Pei}, ``Modulation in the air:
	Backscatter communication over ambient {OFDM} carrier,'' \emph{IEEE Trans.
		Commun.}, vol.~66, no.~3, pp. 1219--1233, 2018.
	
	\bibitem{multi_antenna_Duan_2018}
	R.~{Duan}, R.~{Jantti}, M.~{ElMossallamy}, Z.~{Han}, and M.~{Pan},
	``Multi-antenna receiver for ambient backscatter communication systems,'' in
	\emph{2018 IEEE 19th International Workshop on Signal Processing Advances in
		Wireless Communications (SPAWC)}, 2018, pp. 1--5.
	
	\bibitem{Nocohrent_OFDM_ElMossallamy_2019}
	M.~A. {ElMossallamy}, M.~{Pan}, R.~{J\"antti}, K.~G. {Seddik}, G.~Y. {Li}, and
	Z.~{Han}, ``Noncoherent backscatter communications over ambient {OFDM}
	signals,'' \emph{IEEE Trans. Commun.}, vol.~67, no.~5, pp. 3597--3611, 2019.
	
	\bibitem{ShanpuShen2017_TAP_EHPIXEL}
	S.~Shen, C.~Y. Chiu, and R.~D. Murch, ``Multiport pixel rectenna for ambient
	{RF} energy harvesting,'' \emph{IEEE Trans. Antennas Propag.}, vol.~66,
	no.~2, pp. 644--656, Feb. 2018.
	
	\bibitem{Signal_detection_multiple_antenna_Tag_Kang_2017}
	C.~{Kang}, W.~{Lee}, Y.~{You}, and H.~{Song}, ``Signal detection scheme in
	ambient backscatter system with multiple antennas,'' \emph{IEEE Access},
	vol.~5, pp. 14\,543--14\,547, 2017.
	
	\bibitem{Transceiver_Chen_2020}
	C.~{Chen}, G.~{Wang}, H.~{Guan}, Y.~{Liang}, and C.~{Tellambura}, ``Transceiver
	design and signal detection in backscatter communication systems with
	multiple-antenna tags,'' \emph{IEEE Trans. Wireless Commun.}, vol.~19, no.~5,
	pp. 3273--3288, 2020.
	
	\bibitem{optimal_antenna_Chen_2020}
	C.~{Chen}, G.~{Wang}, P.~D. {Diamantoulakis}, R.~{He}, G.~K. {Karagiannidis},
	and C.~{Tellambura}, ``Signal detection and optimal antenna selection for
	ambient backscatter communications with multi-antenna tags,'' \emph{IEEE
		Trans. Commun.}, vol.~68, no.~1, pp. 466--479, 2020.
	
	\bibitem{Probability_Papoulis_2002}
	A.~{Papoulis} and S.~U. {Pillai}, \emph{{P}robability, {R}andom {V}ariables and
		{S}tochastic {P}rocesses}, 4th~ed., New York, NY, USA: McGraw-Hill, 2002.
	
	\bibitem{Semi_coherent_Qian_2017}
	J.~{Qian}, F.~{Gao}, G.~{Wang}, S.~{Jin}, and H.~{Zhu}, ``Semi-coherent
	detection and performance analysis for ambient backscatter system,''
	\emph{IEEE Trans. Commun.}, vol.~65, no.~12, pp. 5266--5279, 2017.
	
	\bibitem{OSTBC_1999}
	V.~{Tarokh}, H.~{Jafarkhani}, and A.~R. {Calderbank}, ``Space-time block codes
	from orthogonal designs,'' \emph{IEEE Trans. Inf. Theory}, vol.~45, no.~5,
	pp. 1456--1467, 1999.
	
	\bibitem{alamouti_1998}
	S.~M. {Alamouti}, ``A simple transmit diversity technique for wireless
	communications,'' \emph{IEEE J. Sel. Areas Commun.}, vol.~16, no.~8, pp.
	1451--1458, 1998.
	
	\bibitem{goldsmith_2005}
	A.~Goldsmith, \emph{Wireless communications}, Cambridge university press, 2005.
	
	\bibitem{Blind_channel_estimation_Ma_2018}
	S.~{Ma}, G.~{Wang}, R.~{Fan}, and C.~{Tellambura}, ``Blind channel estimation
	for ambient backscatter communication systems,'' \emph{IEEE Commun. Lett.},
	vol.~22, no.~6, pp. 1296--1299, 2018.
	
	\bibitem{Machine_CSI_mA_2018}
	S.~{Ma}, Y.~{Zhu}, G.~{Wang}, and R.~{He}, ``Machine learning aided channel
	estimation for ambient backscatter communication systems,'' in \emph{2018
		IEEE International Conference on Communication Systems (ICCS)}, 2018, pp.
	67--71.
	
	\bibitem{Sparse_CSI_kIM_2018}
	T.~Y. Kim and D.~I. Kim, ``Novel sparse-coded ambient backscatter communication
	for massive {I}o{T} connectivity,'' \emph{Energies}, vol.~11, no. 1780, pp.
	1--25, Jul. 2018.
	
	\bibitem{CSI_massive_Reader_Zhao_2019}
	W.~{Zhao}, G.~{Wang}, S.~{Atapattu}, R.~{He}, and Y.~{Liang}, ``Channel
	estimation for ambient backscatter communication systems with massive-antenna
	reader,'' \emph{IEEE Trans. Veh. Technol.}, vol.~68, no.~8, pp. 8254--8258,
	2019.
	
	\bibitem{multitag_channel_estimation_Mishra_2019}
	D.~{Mishra} and E.~G. {Larsson}, ``Multi-tag backscattering to {MIMO} reader:
	Channel estimation and throughput fairness,'' \emph{IEEE Trans. Wireless
		Commun.}, vol.~18, no.~12, pp. 5584--5599, 2019.
	
	\bibitem{joint_CSI_Darsena_2018}
	D.~{Darsena}, G.~{Gelli}, and F.~{Verde}, ``Joint channel estimation,
	interference cancellation, and data detection for ambient backscatter
	communications,'' in \emph{2018 IEEE 19th International Workshop on Signal
		Processing Advances in Wireless Communications (SPAWC)}, 2018, pp. 1--5.
	
	\bibitem{differential_Tarokh_2000}
	V.~{Tarokh} and H.~{Jafarkhani}, ``A differential detection scheme for transmit
	diversity,'' \emph{IEEE Journal on Selected Areas in Communications},
	vol.~18, no.~7, pp. 1169--1174, 2000.
	
	\bibitem{multiple_diff_STBC_2001}
	H.~{Jafarkhani} and V.~{Tarokh}, ``Multiple transmit antenna differential
	detection from generalized orthogonal designs,'' \emph{IEEE Trans. Inf.
		Theory}, vol.~47, no.~6, pp. 2626--2631, 2001.
	
	\bibitem{2port_1}
	S.~{Shen}, C.~{Chiu}, and R.~D. {Murch}, ``A dual-port triple-band {L}-probe
	microstrip patch rectenna for ambient {RF} energy harvesting,'' \emph{IEEE
		Antennas Wireless Propag. Lett.}, vol.~16, pp. 3071--3074, 2017.
	
	\bibitem{2port_2}
	S.~{Shen} and R.~D. {Murch}, ``Designing dual-port pixel antenna for ambient
	{RF} energy harvesting using genetic algorithm,'' in \emph{2015 IEEE
		International Symposium on Antennas and Propagation USNC/URSI National Radio
		Science Meeting}, 2015, pp. 1286--1287.
	
	\bibitem{ShanpuShen2019_TMTT_Freqdepend}
	S.~{Shen} \emph{et~al.}, ``An ambient {RF} energy harvesting system where the
	number of antenna ports is dependent on frequency,'' \emph{IEEE Trans.
		Microw. Theory Tech.}, vol.~67, no.~9, pp. 3821--3832, Sep. 2019.
	
	\bibitem{ShanpuShen2019_TIE_HybridCombining}
	S.~{Shen}, Y.~{Zhang}, C.~{Chiu}, and R.~{Murch}, ``A triple-band high-gain
	multibeam ambient {RF} energy harvesting system utilizing hybrid combining,''
	\emph{IEEE Trans. Ind. Electron.}, vol.~67, no.~11, pp. 9215--9226, 2020.
	
\end{thebibliography}
\end{document}